\documentclass[aps,pra,amsmath,amssymb,twocolumn,superscriptaddress,showpacs,floatfix,]{revtex4-1}

\usepackage{ifpdf}
\newif\ifpdf
\ifx\pdfoutput\undefined
\pdffalse
\else
\pdfoutput=1
\pdftrue
\fi
\ifpdf
\usepackage{graphicx}
\usepackage{epstopdf}
\else
\usepackage{graphicx}
\fi

\usepackage{bm}
\usepackage{epsfig}
\usepackage[usenames]{color}
\usepackage{soul}
\usepackage{graphicx}
\usepackage{hyperref}
\usepackage{amsmath}

\begin{document}


\title{Mode cooperation in two-dimensional plasmonic distributed-feedback laser}

\author{N.~E.~Nefedkin}
\email{nefedkin@phystech.edu}
\affiliation{Dukhov Research Institute of Automatics (VNIIA), 22 Sushchevskaya, Moscow 127055, Russia}
\affiliation{Moscow Institute of Physics and Technology, Moscow 141700, Russia}
\affiliation{Institute for Theoretical and Applied Electromagnetics, 13 Izhorskaya, Moscow 125412, Russia}

\author{A.A. Zyablovsky}
\affiliation{Dukhov Research Institute of Automatics (VNIIA), 22 Sushchevskaya, Moscow 127055, Russia}
\affiliation{Moscow Institute of Physics and Technology, Moscow 141700, Russia}

\author{E.S. Andrianov}
\affiliation{Dukhov Research Institute of Automatics (VNIIA), 22 Sushchevskaya, Moscow 127055, Russia}
\affiliation{Moscow Institute of Physics and Technology, Moscow 141700, Russia}

\author{A.~A.~Pukhov}
\affiliation{Dukhov Research Institute of Automatics (VNIIA), 22 Sushchevskaya, Moscow 127055, Russia}
\affiliation{Moscow Institute of Physics and Technology, Moscow 141700, Russia}
\affiliation{Institute for Theoretical and Applied Electromagnetics, 13 Izhorskaya, Moscow 125412, Russia}

\author{A.~P.~Vinogradov}
\affiliation{Dukhov Research Institute of Automatics (VNIIA), 22 Sushchevskaya, Moscow 127055, Russia}
\affiliation{Moscow Institute of Physics and Technology, Moscow 141700, Russia}
\affiliation{Institute for Theoretical and Applied Electromagnetics, 13 Izhorskaya, Moscow 125412, Russia}

\date{\today}

\begin{abstract}

Plasmonic distributed-feedback lasers based on a two-dimensional periodic array of metallic nanostructures are the main candidate for nanoscale sources of coherent electromagnetic field.
Strong localization of the electromagnetic field and the large radiation surface are good opportunities for achieving an ultrashort response time to the external actions and creating beam directionality.
At the same time, the investigation of such a  system is a challenging problem.
In this paper, we present an exhaustive study of the operation of a two-dimensional plasmonic distributed-feedback laser.
We show that the complex structure of the modes of a periodic plasmonic array and the nonlinear interaction between the modes through the active medium lead to a new effect, namely, mode cooperation.
Mode cooperation is manifested as the generation of the modes in an allowed band with a high threshold instead of modes localized near the band gap with a low threshold.
Suppression of lasing of the modes at the edge of the band gap results in widening of the radiation pattern above the generation threshold.
This paves the way for effective control and manipulation of the radiation pattern of nanoscale systems, which is of great importance for applications in spectroscopy and optoelectronics. 
\end{abstract}

\maketitle

\section{Introduction}
\label{sec:intro}

The creation of nanosized sources of coherent radiation is a challenging problem in modern laser science~\cite{Ning2016SCLas,Hill2014SmallLas}.
The size of ordinary lasers is constrained by the diffraction limit.
To overcome this, it has been proposed to use metallic nanostructures as laser cavities~\cite{BergmanPRL,StockmanJOpt} with resonance plasmonic excitations as cavity modes.
In the last decade, substantial progress has been achieved in creating and investigating lasers based on plasmonic nanostructures, such as metallic nanoparticles~\cite{noginov2009demonstration,totero2016energy}, waveguides~\cite{hill2009lasing,oulton2009plasmon,mayer2015monolithically,pickering2014cavity,lu2012plasmonic}, etc.
To enhance feedback and obtain beam directionality and a narrow radiation pattern it has been suggested to use periodic metallic structures.
The periodicity leads to the formation of allowed and forbidden bands for the electromagnetic field.
The electromagnetic field in periodic structures  is presented as a superposition of Bloch modes with frequencies belonging to an allowed band and wavenumbers from $0$ to $2 \pi / L$, where $L$ is the system's period~\cite{Fan1997PC,JoannopoulosPC}.
These Bloch modes are distributed throughout the resonator.
The interaction of the Bloch modes with the active medium creates positive distributed feedback~\cite{ZhouNatNano,BeijnumPRL,ExterOptExp,TennerJOpt,TennerACSPhot,MengLPR,SuhNL,SchokkerPRB,SchokkerACSPhot,YangNatComm,hakala2017lasing,ramezani2017plasmon,YangACSNano}.
Such plasmonic distributed-feedback (DFB) lasers have been realized in a series of experiments, where periodic plasmonic structures, such as metallic films perforated by holes~\cite{BeijnumPRL,ExterOptExp,MengLPR,TennerJOpt,TennerACSPhot,TennerOptExp,Melentiev2017nanolaser} or two-dimensional arrays of plasmonic nanoparticles~\cite{SchokkerPRB,SchokkerACSPhot,YangACSNano,ZhouNatNano,hakala2017lasing,schokker2016lasing,YangNatComm,Odom2017nanolasing,Odom2017PlasmonCoherence,Odom2017PlasmonEngin} play the role of the resonator.

Plasmonic DFB lasers are very promising for a wide range of applications, from sensing and bioimaging to optoelectronics~\cite{Hill2014SmallLas}.
For spectroscopic and optoelectronic applications, it is necessary that a laser has a narrow tunable radiation pattern.
The radiation pattern of a plasmonic DFB laser is created by the radiation pattern of the Bloch modes which propagate along the plane of the surface of the DFB laser.
Standard qualitative arguments predict that a DFB laser has to radiate in a narrow cone close to the normal direction.
Indeed, the radiation angle of these modes as well as their radiation losses depend on the ratio of the Bloch wavenumber to the free space wavenumber~\cite{TennerJOpt}.
For eigenmodes at the edge of a band gap, this ratio tends to zero, and such modes radiate close to the normal direction of the DFB laser surface while modes in the allowed band radiate at a nonzero angle.
In the vicinity of the edge of the band gap, there are modes for which the radiation losses are the smallest~\cite{Turnbull2001PCDFB,Kazarinov1985DFB}.
These modes have the lowest generation threshold and have to be generated with the highest amplitude.
This should lead to radiation in a narrow cone close to the normal direction, and prevent the tunability of the radiation pattern.
However, recent experiments~\cite{Odom2017PlasmonCoherence} show much more complex behavior of the radiation pattern, which cannot be explained by these qualitative arguments.

\begin{figure}[b]
	\centering
	\includegraphics[width=0.95\columnwidth]{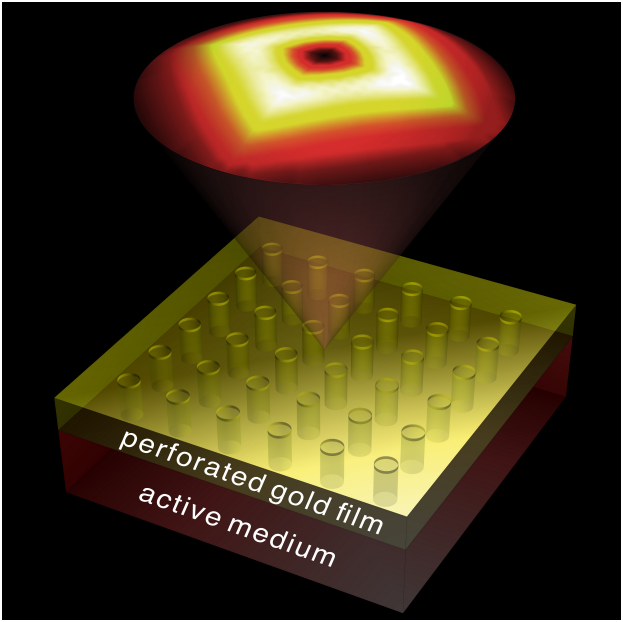}\\
	\caption{A sketch of 2D plasmonic DFB laser consisting of two layers and the radiation pattern created by the laser. The first layer is a gold film perforated by an array of nanoholes and the second layer is an active medium located beneath the gold film.}\label{fig1}
\end{figure}

In the present paper, combining comprehensive numerical simulations with a simple analytical model, we show that the complex structure of the modes of a periodic plasmonic lattice and the nonlinear interaction between modes through the active medium lead to a new effect, called mode cooperation.
This effect appears as a growth of stimulated emission in the modes in the allowed band with a high threshold instead of the modes localized near the band gap with a low threshold.
Mode cooperation is the opposite of the mode competition in usual lasers.
Suppression of the lasing of the modes near the band gap results in the widening of the radiation pattern above the generation threshold.
We show that mode cooperation depends strongly on the density of states of the system.
We also demonstrate that mode cooperation takes place for two-dimensional DFB lasers and does not appear in one-dimensional ones.
Mode cooperation gives the opportunity for effective control and manipulation of the radiation pattern of nanolaser systems, which can be widely applied in spectroscopy and optoelectronics.

\section{The model for describing the dynamics of a plasmonic DFB laser}
\label{sec:markupcmd}
The description of the dynamics of the electromagnetic field of a periodic plasmonic lattice and the atoms of the active medium (see Fig.~\ref{fig1}) is based on the approach which has been proposed in~\cite{Zyablovsky2017approach} for considering ultrafast phenomena in dispersive dissipative media.
The basis of this approach is the description of the collective dynamics of the photon number in each mode, the population inversion of each atom, and the energy flows between the different modes through the active medium.
For the system under consideration, this approach leads to the following equations (see Note 1, 4 in Suppl. Mat.)
\begin{gather}\label{MainEqs}
\frac{{d{n_{jj}}}}{{dt}} =  - 2{\gamma _j}{n_{jj}} + \sum\limits_m {\frac{{{{\left| {{\Omega _{jm}}} \right|}^2}{\gamma _{\sigma j}}}}{{{{\left| {{\Delta _j}} \right|}^2}}}} \left( {2{n_{jj}}{D_m} + {D_m} + 1} \right) + 
\\
\nonumber +\sum\limits_{m,k,k \ne j} {\left( {\frac{{{\Omega _{jm}}\Omega _{km}^*{n_{jk}}}}{{{\Delta _j}}} + {\rm{c}}{\rm{.c}}{\rm{.}}} \right){D_m}}
\\
\frac{{d{n_{jl}}}}{{dt}} =  - \left( {{\gamma _j} + {\gamma _l}} \right){n_{jl}} + i\left( {{\omega _j} - {\omega _l}} \right){n_{jl}} + 
\\
\nonumber +\sum\limits_m {\left( {\frac{{{\Omega _{lm}}\Omega _{jm}^*\left( {{D_m} + 1} \right)\left( {{\Delta _j} + \Delta _l^*} \right)}}{{2{\Delta _j}\Delta _l^*}}} \right)}  +
\\
\nonumber + \sum\limits_{m,k,k \ne j} {\left( {\frac{{{\Omega _{lm}}\Omega _{km}^*{n_{jk}}}}{{{\Delta _j}}} + \frac{{{\Omega _{km}}\Omega _{jm}^*{n_{lk}}}}{{\Delta _l^*}}} \right){D_m}} 
\\
\frac{{d{D_m}}}{{dt}} =  - {\gamma _D}\left( {1 + {D_m}} \right) + {\gamma _{pump}}\left( {1 - {D_m}} \right) - 
\\
\nonumber -2\sum\limits_j {\frac{{{{\left| {{\Omega _{jm}}} \right|}^2}{\gamma _{\sigma j}}}}{{{{\left| {{\Delta _j}} \right|}^2}}}} \left( {2{n_{jj}}{D_m} + {D_m} + 1} \right) - 
\\
\nonumber -2\sum\limits_{j,k,k \ne j} {\left( {\frac{{{\Omega _{jm}}\Omega _{km}^*{n_{jk}}}}{{{\Delta _j}}} + {\rm{c}}{\rm{.c}}{\rm{.}}} \right){D_m}} 
\end{gather}
Here, ${n_{jj}}$ is the photon number in the $j$th mode, and ${n_{jl}}$ is an interference term responsible for the flow of photons from the $l$th cavity mode to the $j$th cavity mode when $j \ne l$.
${D_m}$ is the mean value of the population inversion of the $m$th atom.
The parameters $\omega_i$, $\gamma_i$ correspond to the real and imaginary parts of the modes' eigenfrequencies (see Note 2 in Suppl. Mat.); $\omega_\sigma$, $\gamma_{ph}$, $\gamma_D$ $\gamma_{pump}$ are the atom transition frequency and atom rates of energy relaxation, phase relaxation, and pumping, respectively.
${\Delta _j} = {\gamma _{\sigma j}} - i\left( {{\omega _j} - {\omega _\sigma }} \right)$, where ${\gamma _{\sigma j}} = {\gamma _{ph}} + {\gamma _j} + \left( {{\gamma _D} + {\gamma _{pump}}} \right)/2$.
$\Omega_{jm}$ is the Rabi constant of interaction between the $j$th mode and the $m$th atom (See Note 3 in Suppl. Mat.).
These equations generalize the rate equations widely used in laser theory.
In addition to the dynamics of the photon number in each mode and the population inversion of each atom, they describe additional energy flows between the active medium and the atoms.
Such energy flows arise from the stimulated emission generated by interference terms, $n_{jl}$.

The energy flows are described by the last terms in Eqs.~(\ref{MainEqs})--(3) and strongly depend on the spatial overlap of the plasmonic lattice modes in the active medium region.
Indeed, the last terms in Eqs.~(\ref{MainEqs})--(3) are proportional to 
\begin{gather}\label{MainEqs0}
\begin{array}{l}
\sum\limits_{m,k,k \ne j} {\left( {\frac{{{\Omega _{jm}}\Omega _{km}^*{n_{jk}}}}{{{\Delta _j}}} + c.c.} \right)} {D_m} = \\
= \sum\limits_{k,k \ne j} {\frac{{{n_{jk}}}}{{{\Delta _j}}}\sum\limits_m {\left( {{\Omega _{jm}}\Omega _{km}^*{D_m} + c.c.} \right)} }  \sim \\
\sim \sum\limits_{k,k \ne j} {\frac{{{n_{jk}}\bar D}}{{{\Delta _j}}}\int\limits_{{V_{gain}}} {\left( {E_k^*\left( {\bf{r}} \right){E_j}\left( {\bf{r}} \right) + c.c.} \right)dV} } 
\end{array}
\end{gather}
  where the range of integration coincides with the pump area, $\bar D$  is the characteristic value of the population inversion in the active medium, and the $n_{jk}$ play role of the interference terms of the electromagnetic field~\cite{Zyablovsky2017approach}.
 When the active medium is pumped uniformly over the whole volume, these energy flows are equal to zero due to the orthogonality of the modes.
 In such a case, the best lasing condition is for the mode with the smallest dissipation and the highest constant of interaction with the active medium~\cite{HakenLaser}.
 However, when only part of the active medium is pumped, the interference terms are nonzero and energy flows become essential.
 Such energy flows take their maximum values for the modes which are maximally overlapped in the region where the active medium is located.
 As we show below, these terms are crucial for two-dimensional plasmonic lasers with inhomogeneous pumping.

\begin{figure*}
	\centering
	\includegraphics[width=1.95\columnwidth]{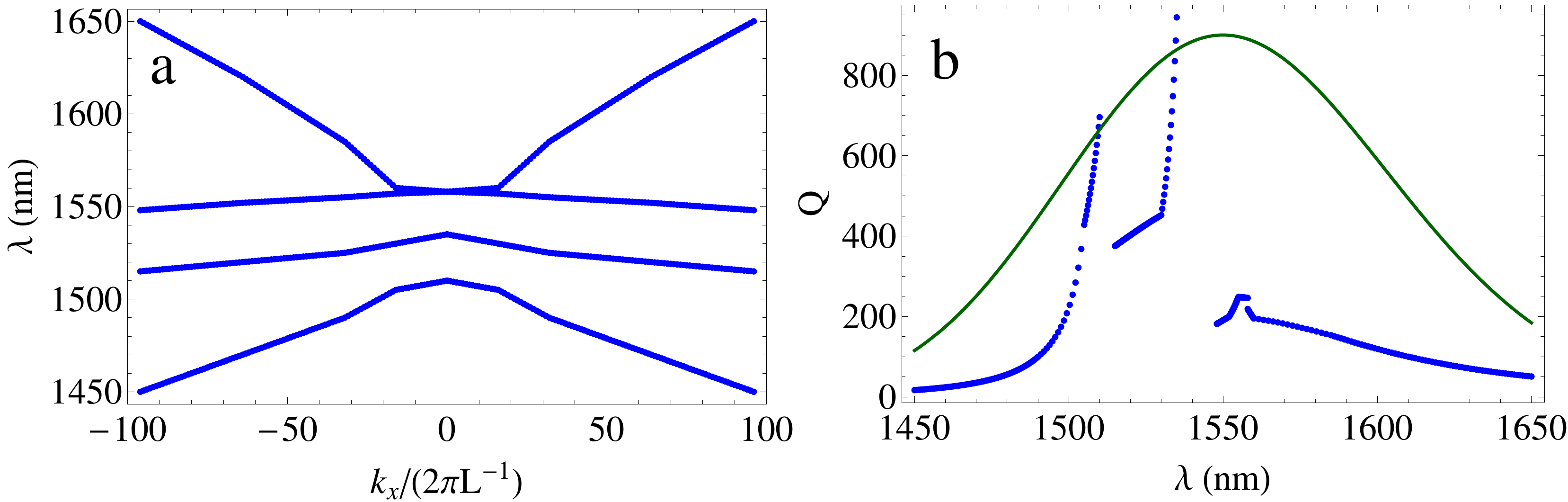}\\
	\caption{(a) The dispersion curve of the 2D plasmonic DFB laser. The surface area of Au film is $L \times L = 190 \times 190\,\mu {m^2}$. The step of quantization of Bloch wavenumber $\delta {k_B} = \pi /L$. (b) The quality factor of eigenmodes. The green line is an atomic emission line.}\label{fig2}
\end{figure*}

\begin{figure}
	\centering
	\includegraphics[width=0.95\columnwidth]{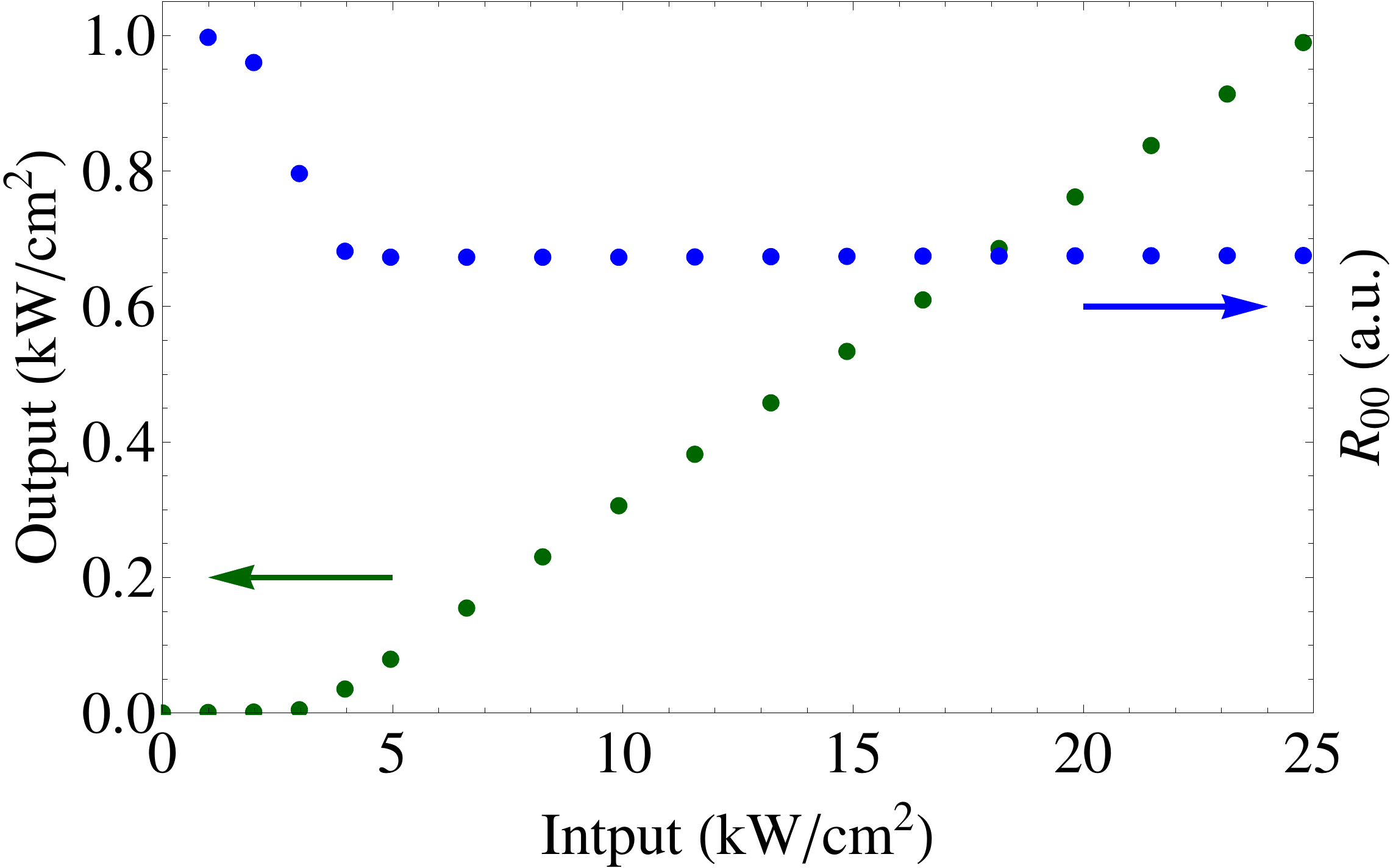}\\
	\caption{The dependence of output power (green points) and the ratio of the photon number in the mode at the edge of a band gap to the total photon number in all modes ${R_{00}} = n\left( {{k_{Bx}} = 0,{k_{By}} = 0} \right)/\sum\limits_i {{n_{ii}}}$ (blue points) on the input power.}\label{fig3}
\end{figure}

\section{Results and discussion}
\subsection{Lasing curve and the radiation pattern of two-dimensional plasmonic DFB laser}
We now consider the 2D plasmonic DFB laser with cavity based on Au film perforated by a periodic array of nanosized holes, see Fig.~\ref{fig1}.
To find the eigenmodes of the plasmonic periodic nanostructure, we used the coupled-mode theory with the experimental values of the scattering rates (see~\cite{TennerJOpt} and Note 2 in Suppl. Mat.).
The dispersion curve of this system is shown in Fig.~\ref{fig2}a.
There are several band gaps in investigated wavelength range.
Waves with a wavelength belonging to band gaps cannot propagate along the structure.
At the edge of a band gap, there are dark modes with Bloch wavevector $k_B = (0,0)$.
Such modes do not radiate and, thus, have the lowest dissipation rates.
Other modes in allowed band with $k \ne 0$ radiate at angle $\sin \theta = k/(2 \pi / \lambda_0)$, where $\lambda_0 = 2 \pi c / \omega$.
The active medium is located beneath the surface of the metallic film.
We suppose that the width of the active medium transition line covers whole investigated frequency range, see Fig \ref{fig2}b.
This condition is typical for experiments~\cite{TennerJOpt,TennerACSPhot}.

 \begin{figure*}[t]
 	\centering
 	\includegraphics[width=1.95\columnwidth]{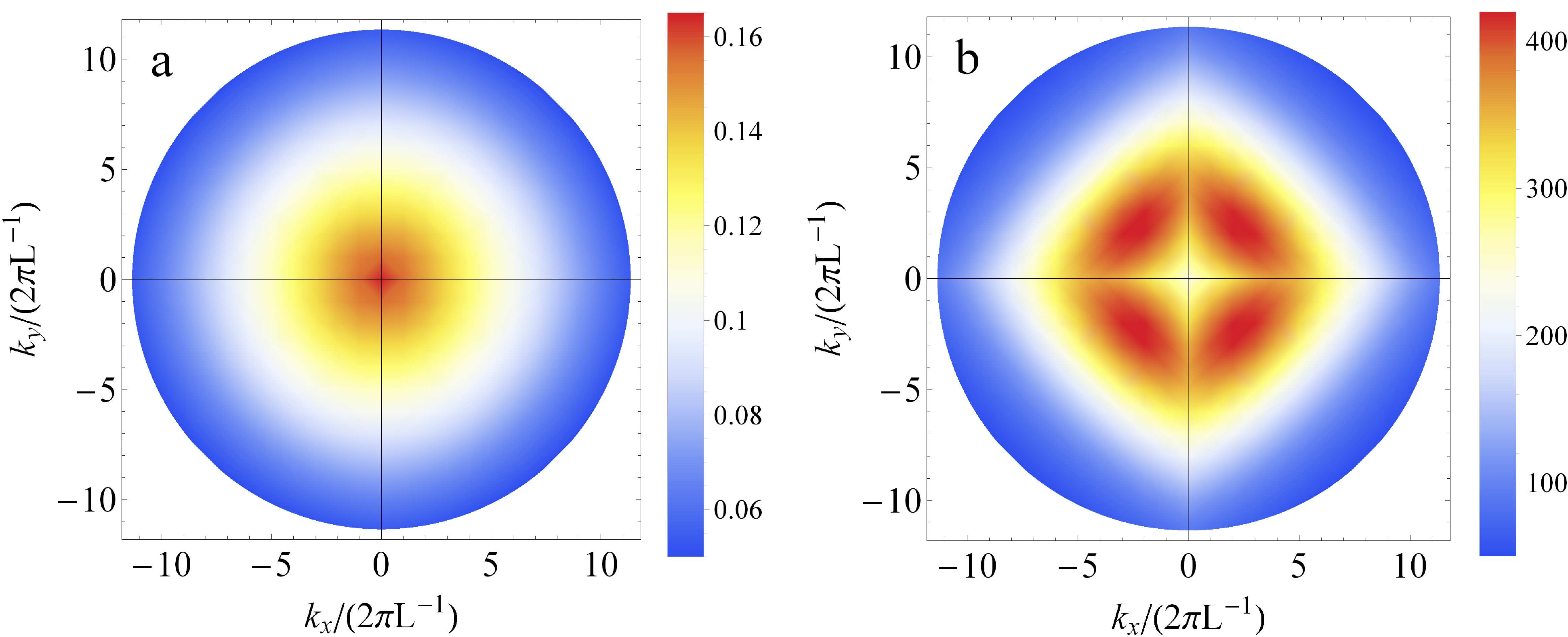}\\
 	\caption{The mode distribution of the photon number below (a) and above (b) the pump threshold. }\label{fig4}
 \end{figure*}  
  
We consider the case when the pump beam is focused to a finite spot on the plasmonic laser surface~\cite{ZhouNatNano,BeijnumPRL,ExterOptExp,TennerJOpt,TennerACSPhot,MengLPR,SuhNL,SchokkerPRB,SchokkerACSPhot,YangNatComm,hakala2017lasing,ramezani2017plasmon,YangACSNano}.
In the pump area, the population inversion is positive, whereas beyond the pump area it remains negative~\cite{ZhouNatNano}.
The electromagnetic field, in turn, is amplified inside the pump area until its amplitude reaches the stationary value.
However, beyond the pump area amplitudes of the electromagnetic waves exponentially decrease at a distance from active region.
This structure is similar to one investigated in the papers~\cite{BeijnumPRL,TennerJOpt,TennerACSPhot,MengLPR}.
The mode competition in a multimode laser usually leads to suppression of lasing in modes with a high pump threshold~\cite{oulton2009plasmon,Stone2008RandomLaser,Zyablovsky2017Optimum}.
For this reason, we can expect that the lasing could be observed in dark modes.
The photon number in these modes would be much higher than in other modes.
In this case, the radiation of the DFB laser is rather small.

The main results of numerical simulations of Eqs. (\ref{MainEqs})--(3) are shown in Figs.~\ref{fig3}--\ref{fig4}.
The dependence of the total output electromagnetic field intensity on the input pump power is shown in Fig.~\ref{fig3}.
It is seen that the output intensity demonstrates threshold behavior, see green points in Fig.~\ref{fig3}.
The distribution of the photon number below and above the pumping threshold over the modes with different values of the Bloch wavevector ${{\bf{k}}_B} = \left( {k_x,k_y} \right)$  is shown in Figs.~\ref{fig4} (a) and (b), respectively.
On one hand, below the threshold the photon number is maximal for the dark mode at the edge of a band gap with low dissipation rates, see Fig.~\ref{fig4}a.
The photon number monotonically decreases when the mode frequency moves from the edge of the band gap to the allowed band.
In other words, the excitation is maximal for the dark modes with ${{\bf{k}}_B} = \left( {0,0} \right)$.
This result is expected from the qualitative considerations.

On the other hand, the situation dramatically changes above the threshold.
The maximum of the photon number is observed for the modes which lie in the allowed band, see Fig.~\ref{fig4}b.
Such behavior contradicts the intuitive expectations.
Usually, the modes with the lowest threshold have a maximal field amplitude.
However, in investigated two-dimensional plasmonic DFB laser the modes, which lie inside the allowed band and have higher dissipation rates due to radiation, have maximal amplitudes.
Thus, stimulated emission is primarily in bright modes with ${{\bf{k}}_B} \ne \left( {0,0} \right)$.

In Fig.~\ref{fig3}, we also show the ratio of the photon number in the mode at the edge of the band gap to the total photon number in all modes (blue points in Fig.~\ref{fig3}).
It is seen that this ratio decreases from maximal value below the threshold to a constant value above the threshold.
This corresponds to the fact that above the threshold an essential part of the total output intensity is produced by the bright modes.
Transition from spontaneous emission in dark modes to lasing in bright ones leads to both a sharp increase of the photon number and the radiation power.

Thus, below the generation threshold, when the main contribution in laser output is produced by spontaneous transitions of atoms of the active medium, the photon number is maximal in the dark mode at the edge of a band gap.
At the same time, above the threshold, when the stimulated emission dominates, the maximum photon number is in the bright modes lying inside the allowed band.

\subsection{The mode cooperation}
The described situation is unusual for lasers.
Conventionally the modes with the lowest generation threshold has a maximum photon number.
In the previous section, we have seen that this behavior takes place only under the generation threshold.
The reason can be revealed through a simple system which consists of three modes with eigenfrequencies  ${\omega _2} = {\omega _3}$, $\left| {{\omega _1} - {\omega _2}} \right| \gg {\gamma _2}$ and dissipation rates ${\gamma _1} < {\gamma _2} = {\gamma _3}$ which interact with the active medium.
The second and third modes differ in the field distribution.
For simplicity we suppose that ${\gamma _{ph}} \gg {\gamma _j},{\gamma _D},{\gamma _{pump}},\left| {{\omega _1} - {\omega _2}} \right|$ and the population inversions of all atoms are the same.
In such a case, Eqs. (\ref{MainEqs})--(3) can be simplified and take the form
\begin{gather} \label{eq5}
\frac{{d{n_{11}}}}{{dt}} =  - 2{\gamma _1}{n_{11}} + \frac{\Gamma^2 }{{{\gamma _{ph} }}}\left( {D + 1} \right) + \frac{{2\Gamma^2 }}{{{\gamma _{ph} }}}{n_{11}}D
\\
\frac{{d{n_{22}}}}{{dt}} =  - 2{\gamma _2}{n_{22}} + \frac{\Gamma^2 }{{{\gamma _{ph} }}}\left( {D + 1} \right) +
\\
\nonumber + \frac{{2\Gamma^2 }}{{{\gamma _{ph} }}}{n_{22}}D + \frac{{2{\Gamma^2 _{23}}}}{{{\gamma _{ph} }}}{n_{23}}D  
\\
\frac{{d{n_{23}}}}{{dt}} =  - \left( {{\gamma _2} + {\gamma _3}} \right){n_{23}} + \frac{{{\Gamma^2_{23}}}}{{{\gamma _{ph} }}}\left( {D + 1} \right) + 
\\
\nonumber + \frac{{2\Gamma^2 }}{{{\gamma _{ph} }}}{n_{23}}D + \frac{{2{\Gamma^2 _{23}}}}{{{\gamma _{ph} }}}{n_{22}}D
\\
\frac{{dD}}{{dt}} =  - {\gamma _D}\left( {1 + D} \right) + {\gamma _{pump}}\left( {1 - D} \right) - \frac{{6{\Gamma ^2}}}{{{N_{at}}{\gamma _{ph}}}}\left( {D + 1} \right) -
\\
\nonumber  -\frac{{2D}}{{{N_{at}}}}\left( {\frac{{2{\Gamma ^2}}}{{{\gamma _{ph}}}}\left( {{n_{11}} + {n_{22}} + {n_{33}}} \right)} \right) - \frac{{2D}}{{{N_{at}}}}\left( {\frac{{2\Gamma _{23}^2{n_{23}}}}{{{\gamma _{ph}}}} + {\rm{c}}.{\rm{c}}.} \right)
\end{gather}
where $\Gamma^2 = \sum\limits_{m = 1}^{{N_{at}}} {\Omega _{jm}^*{\Omega _{jm}}} $ is a net interaction constant of the active medium with field of these three modes and $\Gamma _{23}^2 =\sum\limits_{m = 1}^{{N_{at}}} {\Omega _{2m}^*{\Omega _{3m}}} $  is an overlap integral of the second and third modes, and $N_{at}$ is the number of atoms in the system.
Here we neglect the interference terms between the first and other modes (i.e.
the terms ${n_{12}}$ , ${n_{13}}$ ) due to $\left| {{\omega _1} - {\omega _2}} \right| \gg {\gamma _2}$.

In the steady-state, the photon number in the first and second modes is given by
\begin{gather}\label{eq6}
{n_{11}^{st}} = \frac{1}{2}\frac{\Gamma^2 }{{{\gamma _1}{\gamma _{ph} }}}\frac{{D^{st} + 1}}{{1 - \frac{\Gamma^2 }{{{\gamma _1}{\gamma _{ph} }}}D^{st}}}
\\
{n_{22}^{st}} ={n_{33}^{st}} = \frac{1}{2}\frac{{{\Gamma^2 _{eff}}}}{{{\gamma _2}{\gamma _{ph} }}}\frac{{D^{st} + 1}}{{1 - \frac{{{\Gamma^2 _{eff}}}}{{{\gamma _2}{\gamma _{ph} }}}D^{st}}}
\end{gather}
where
\begin{gather}\label{eq7}
{\Gamma^2_{eff}} = \Gamma^2  + \frac{{\Gamma _{23}^4}}{{{\gamma _2}{\gamma _{ph} }}}\frac{D^{st}}{{1 - \frac{\Gamma^2 }{{{\gamma _2}{\gamma _{ph} }}}D^{st}}}
\end{gather}
is an effective coupling-constant between the second mode and the active medium.

\begin{figure}[t]
	\centering
	\includegraphics[width=0.95\columnwidth]{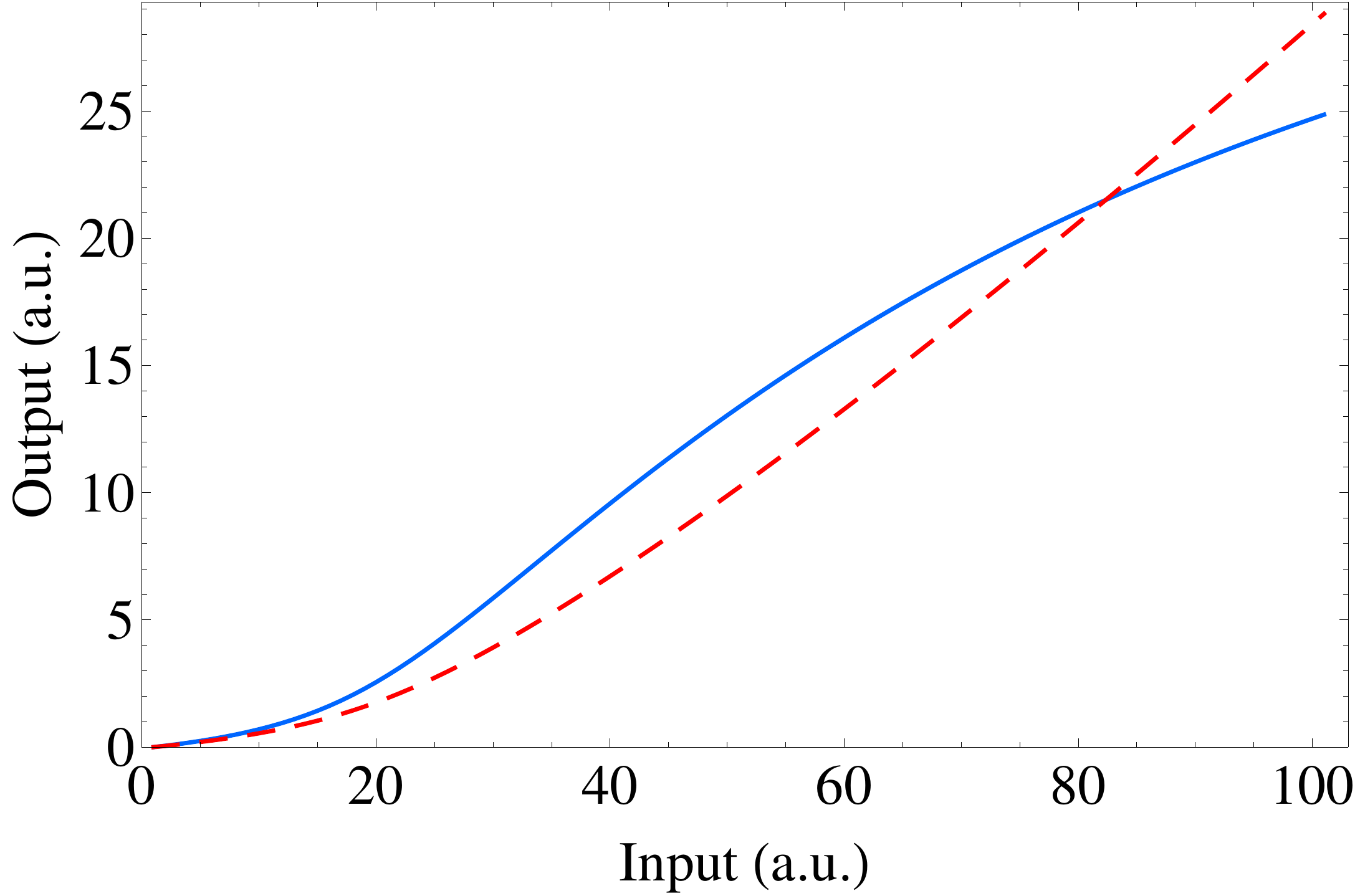}\\
	\caption{The dependence of the photon number in the first (blue solid line) and second (red dashed line) modes on pump intensity calculating from Eqs. (\ref{eq5}) -- (8). Here ${\gamma _1} = {10^{ - 4}}{\omega _\sigma }$; ${\gamma _{2,3}} = 1.5 \times {10^{ - 4}}{\omega _\sigma }$; ${\gamma _D} = {10^{ - 6}}{\omega _\sigma }$; ${\Gamma ^2}/{\gamma _{ph} } = {10^{ - 3}}{\omega _\sigma }$; $\Gamma _{23}^2/{\gamma _{ph} } = 7 \times {10^{ - 4}}{\omega _\sigma }$  and ${N_{at}} = {10^4}$, where ${\omega _\sigma }$ is a transition frequency of atoms of the active medium.}\label{fig5}
\end{figure}

An increase of the pumping rate $\gamma_{pump}$ leads to the rise of the population inversion $D$, see Eq. (8).
This results in a growth of the effective coupling-constant, ${\Gamma _{eff}}$.
From Eqs. (\ref{eq6})--(10) it is seen that when $ {\Gamma^2 _{eff}}/{\gamma _2} < \Gamma^2 /{\gamma _1} $  the photon number in the first mode is larger than the photon number in the second and third modes, see Fig.~\ref{fig5}.
However, when the pump power exceeds the critical value such that $ {\Gamma^2 _{eff}}/{\gamma _2} > \Gamma^2 /{\gamma _1} $, the photon number in the first mode is less than one in the second and third modes, see Fig.~\ref{fig5}.
This behavior is caused by the constructive interference of the second and third modes.
In Eqs. (\ref{eq5})--(8) it reveals as existence of constant ${\Gamma_{23}}$ which describes mode overlap in the active region.
Indeed, when $\Gamma _{23}=0$ the effective coupling-constant ${\Gamma _{eff}}=\Gamma$ and the photon number in each mode is determined by the mode dissipation rate, see denominator in Eqs.~(\ref{eq6})--(10).
However, when $\Gamma_{23} \ne 0$ the effective coupling constant $\Gamma_{eff}$ is increased by the growth of the pumping rate.

Thus, the constructive interference leads to the following consequences.
The interaction between the active medium and the modes with the high threshold (the second and third modes in above example)  is stronger than interaction between the active medium and the mode with the low threshold (the first mode in above example).
As a result, the photon number is the greatest in the modes with high threshold.
Enhancement of the electromagnetic field amplitude due to the constructive interference between modes we call the mode cooperation.

\begin{figure}[t]
	\centering
	\includegraphics[width=0.95\columnwidth]{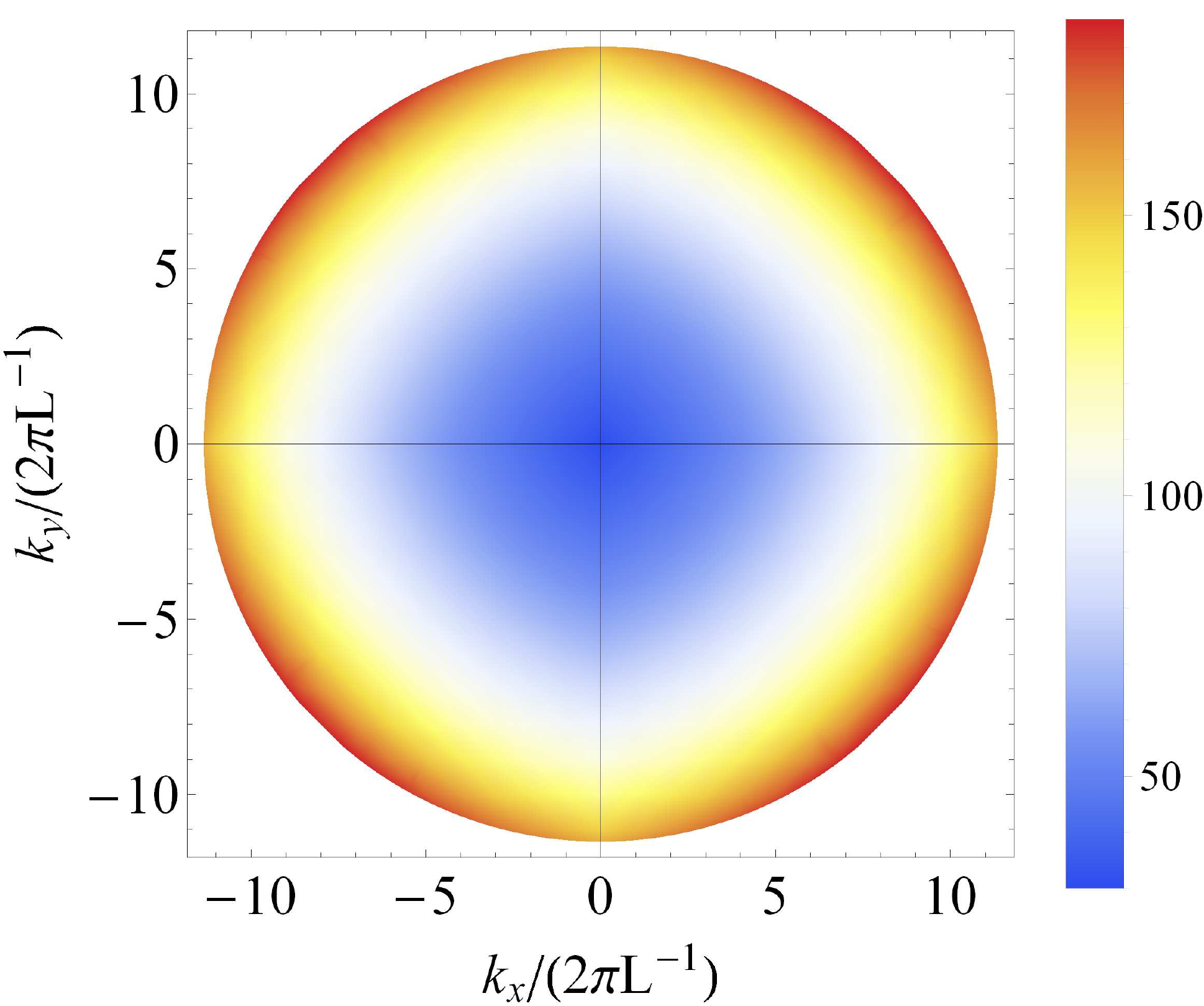}\\
	\caption{The ratio of the energy flow from all atoms to the jth mode with and without the interference terms, i.e. $\left| {\sum\limits_m {\sum\limits_l {{\left| {{\Omega _{jm}}} \right|^2}{n_{jl}}{D_m}} } } \right| / \left| {\sum\limits_m {{\left| {{\Omega _{jm}}} \right|^2}{n_{jj}}{D_m}} } \right|$.}\label{fig6}
\end{figure}

To estimate the influence of interference interaction on lasing in the two-dimensional plasmonic DFB laser let us compare the direct stimulated emission rate into the $j$th mode, $\left| {\sum\limits_m {{{\left| {{\Omega _{jm}}} \right|}^2}{n_{jj}}{D_m}} } \right|$, and the total stimulated emission rate into the $j$th mode, $\left| {\sum\limits_{m,l} {{{\left| {{\Omega _{jm}}} \right|}^2}{n_{jl}}{D_m}} } \right|$.
The first value corresponds to photon amplification in the $j$th mode only due to stimulated emission induced by the photons in this mode.
The second one, in addition, takes into account photon amplification in the $j$th mode by stimulated emission induced by the photons in other modes.
We show the value $\left| {\sum\limits_m {\sum\limits_l {{\left| {{\Omega _{jm}}} \right|^2}{n_{jl}}{D_m}} } } \right| / \left| {\sum\limits_m {{\left| {{\Omega _{jm}}} \right|^2}{n_{jj}}{D_m}} } \right|$ in Fig.~\ref{fig6}.
It is seen that the influence of interference interaction is maximal for modes shifted from the edge of a band gap.
The energy flow from atoms of the active medium to the these modes is increased by interference interaction between modes.
Such increase leads to the shift of the lasing peak away from the edge of the band gap.
Thus, the cavity modes help one another to extract energy from the active medium.
This effect is opposite to the mode competition in conventional lasers, and therefore, justifies its name by the mode cooperation.

Let us estimate at which conditions the mode cooperation can take place in the two-dimensional plasmonic DFB laser.
As it was discussed above, the reason for the mode cooperation is interference interaction between modes.
In the Eqs. (\ref{MainEqs})--(3) the term which is responsible for this interference is a spatial overlap of modes in the region of the pumped active medium.
If the population inversion is equal for each atom, then the spatial overlap of the modes in the region of the active medium can be expressed as
\begin{gather}\label{eq8}
\left( {\sum\limits_{k,k \ne j} {\frac{{{\Gamma^2_{jk}}{n_{jk}}}}{{{\gamma _\sigma }}} + \rm{c.c.}} } \right)\bar D
\end{gather}
where ${\Gamma^2_{jk}} = \sum\limits_m {{\Omega _{jm}}\Omega _{km}^*} $.
In the case, when the pump area occupies all surface of the DFB laser, ${\Gamma_{jk}} = 0$ because the cavity modes are orthogonal in the active area.
The interference flows between modes are mutually cancelled.
In a two-dimensional plasmonic DFB laser, usually the pump area is less than the surface of DFB laser and the modes are nonorthogonal in the volume of pumping.
In this case, the interference interaction between modes is not zero and leads to creating an additional flow from the atoms to the modes.

Let us determine which modes give the main contribution to (\ref{eq8}).
In order to do this, we note that the inteference interaction between the $j$th and $k$th modes, Eq. (\ref{eq8}), is comparable with the direct energy flow from the active medium, $\bar D\frac{{{\Gamma^2_{jj}}{n_{jj}}}}{{{\gamma _\sigma }}}$, when, firstly, $\left| {{\Gamma _{jk}}} \right| \sim {\Gamma _{jj}}$ and, secondly, $\left| {{n_{jk}}} \right| \sim {n_{jj}}$.
The first condition is fulfiled when the $j$th and $k$th modes lie in the same allowed band and their Bloch wavevectors satisfy the following inequality (see the integrand in Eq. (4)):
\begin{gather}\label{eq9}
	\left| {{{\vec k}_{Bj}} - {{\vec k}_{Bk}}} \right|l \le 1
\end{gather}
where $l$ is the size of the pump area.
The absolute value of the interference term between the $j$th and $k$th modes  is proportional to $\left| {{n_{jk}}} \right| \sim {\left( {\left( {{\gamma _j} + {\gamma _l}} \right) - i\left( {{\omega _j} - {\omega _l}} \right)} \right)^{ - 1}}$ while $\left| {{n_{jj}}} \right| \sim 2{ {\left( {{\gamma _j} } \right) } ^{ - 1}}$, see Eqs.~(\ref{MainEqs})--(3).
Thus, the second condition is fulfiled when
\begin{gather}\label{eq10}
\left| {{\omega _j} - {\omega _k}} \right| \le {\gamma _j} + {\gamma _k}
\end{gather}

\begin{figure*}[t]
	\centering
	\includegraphics[width=1.95\columnwidth]{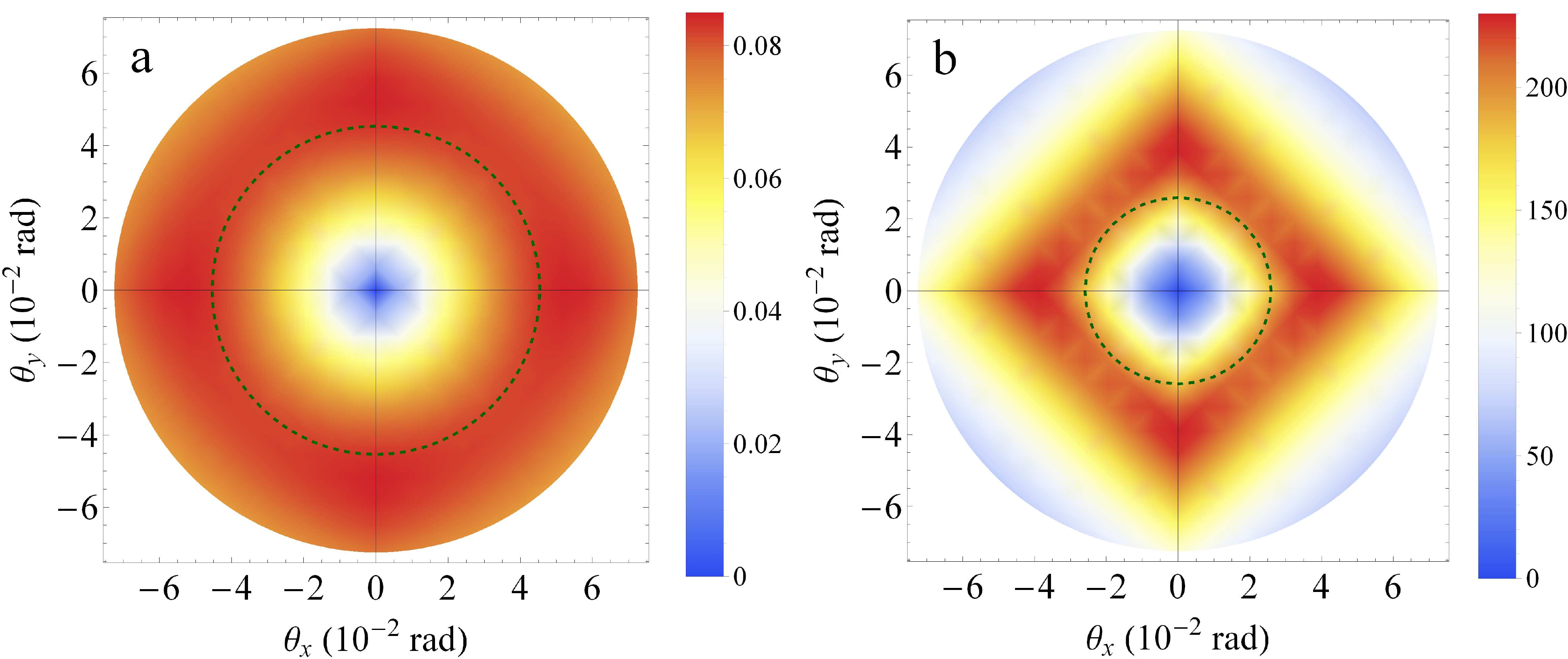}\\
	\caption{The radiation pattern of a 2D plasmonic DFB laser below (a) and above (b) the generation threshold. The dashed lines correspond to the radiation pattern of a one-dimensional plasmonic DFB laser with the same parameters.}\label{fig7}
\end{figure*}

\subsection{Comparison of one- and two-dimensional systems}

The condition (\ref{eq9}) is satisfied for all modes in considered range of wavelength. At the same time, the condition (\ref{eq10}) enables to distinguish the main feature of the radiation of the one- and two-dimensional plasmonic DFB laser. Indeed, the number of modes which lie in the interval $\left( {{\omega _j} - \left( {{\gamma _j} + {\gamma _l}} \right),{\omega _j} + \left( {{\gamma _j} + {\gamma _l}} \right)} \right)$  may be evaluated as $\rho \left( {{\omega _j}} \right)\left( {{\gamma _j} + {\gamma _l}} \right)$, where $\rho (\omega)$ is the density of states. In the two-dimensional case, the density of states $\rho \left( {{\omega _j}} \right) = {\left. {\frac{{{d^2}k}}{{d\omega }}} \right|_{\omega  = {\omega _j}}} \sim \frac{{{k_{Bj}}}}{{{{\left( {\pi /L} \right)}^2}}}{\left. {\frac{{dk}}{{d\omega }}} \right|_{\omega  = {\omega _j}}}$. At the edge of a band gap, $d\omega /dk \sim k_{Bj}$ and, as a consequence, $\rho \left( {{\omega _j}} \right) \sim \rm{const}$ ~\cite{john1994bandgap}. At a distance from the edge of a band gap, the ratio $dk/d\omega$ slightly varies and the density of states is proportional to $k_{Bj}$, see Fig. \ref{fig2}a. Thus, the density of states tends to a constant at the edge of a band gap and increases in an allowed band. As a result, the number of modes in the allowed band which satisfies the condition (\ref{eq10}) is larger than one at the edge of the band gap. So the interference interaction between modes is stronger in the allowed band. Taking into account that the dissipation rate of the modes increases in the allowed band, we come to the conclusion that there is the value of frequency range in the allowed band for which the generation condition is optimal. Numerical simulations confirm this qualitative conclusion, see Fig. \ref{fig4}(a)-(b).

In the one-dimensional case, $\rho \left( {{\omega _j}} \right) \sim  \left( {L/\pi} \right)dk/d\omega $ and has a maximum at the edge of the band gap, where $d{\omega}/dk$ tends to zero ~\cite{john1994bandgap}. As a result, the dark modes at the edge of the band gap have the optimal condition for generation. Numerical simulations of Eqs. (\ref{MainEqs}) -- (3) for the one-dimensional plasmonic DFB laser confirm above conclusion. The maximum of the photon number is observed in the dark mode at the edge of the band gap both below and above the generation threshold (see Note 5, Fig. 1 in Suppl. Mat.)

Thus, in the one-dimensional DFB lasers the dark modes at the edge of the band gap are generated while in the two-dimensional DFB lasers there are bright modes in the allowed band.

\subsection{The radiation pattern of the two-dimensional plasmonic DFB laser}
The existence of mode cooperation can be experimentally verified by measuring the radiation pattern.
As has been noted in the Introduction, the direction of the radiation depends on the ratio of the Bloch wavevector to the wavevector in free space.
Namely, the angle $\theta _{x,y}$ at which the mode of the plasmonic DFB laser radiates is determined by the equations 

\begin{gather}\label{eq11}
{k_{Bx,By}} = \left( {2\pi /\lambda } \right)\sin {\theta _{x,y}}
\end{gather}

On the one hand, if the modes at the edge of the band gap with zero Bloch wavevector are generated, then the radiation pattern represents a narrow cone around the normal direction.
This takes place in the one-dimensional case, when the generation takes place at the edge of the band gap, see the green dashed curves in Fig.~\ref{fig7}(b).
On the other hand, if the modes in an allowed band with non-zero Bloch vector are generated, then the radiation pattern has a maximum in a direction which is determined by the mode cooperation.
Such a situation arises above the generation threshold in the two-dimensional case, see Fig.~\ref{fig7}(b).

In recent experiments on plasmonic lasers~\cite{TennerACSPhot}, it has been shown that the radiation pattern of the two-dimensional plasmonic DFB laser is broadened with respect to the prediction of standard Kogelnik theory~\cite{TennerACSPhot}.
In this theory, the laser modes, their eigenfrequencies, and generation thresholds for the one-dimensional DFB laser are calculated by the coupled-mode theory~\cite{Kogelnik1972DFB}.
The radiation pattern of the DFB laser in such a model coincides with the radiation pattern of the mode with the lowest generation threshold because the given model does not consider the interference interaction between modes through the active medium.
As a result, the predicted radiation angle does not coincide with experiments for the two-dimensional system.

Taking into account the mode interaction, it is shown that in the two-dimensional plasmonic DFB lasers one can observe mode cooperation.
The influence of the mode cooperation explains the discrepancy between the radiation patterns calculated from the standard Kogelnik theory~\cite{Kogelnik1972DFB} and those measured in~\cite{TennerACSPhot}.
Indeed, the mode cooperation results in changing the photon number distribution in the modes above the threshold such that the highest photon number is in the bright modes with nonzero Bloch wavevectors in the plane of the DFB laser (see the blue points in Fig.~\ref{fig3}b).
This leads to a broadening of the radiation pattern, see Fig.~\ref{fig7}(b).
In our calculations, we use the same system parameters as in~\cite{TennerACSPhot}.
A comparison of Fig.~\ref{fig7}(b) in our paper and Fig.~3c in~\cite{TennerACSPhot} demonstrates the good agreement between our theory and the experimental data.

\section{Discussion and Conclusions}

We have presented numerical simulations and analytical analysis of the dynamics of a two-dimensional plasmonic DFB laser.
We found the  photon number, the radiation pattern, and the laser output on the pump intensities.
We have shown that the complex electromagnetic structure and nonlinear interaction through the active medium leads to a new effect, namely, mode cooperation.

Two observed effects follow from mode cooperation.
The first one is a dramatic increase in the laser radiation by crossing the pump threshold.
This is explained by the fact that the spontaneous emission is primarily emitted into the dark modes below the threshold while above the threshold, due to the mode cooperation, the stimulated emission is mainly radiated into the bright modes.
The radiation output in such lasers has a clear threshold.
The second effect is a widening of the radiation pattern.
This is due to the dependence of the radiation pattern on the Bloch wavevector of the mode.
The dark mode at the edge of a band gap has zero Bloch wavevector and does not radiate.
The neighboring modes radiate within a narrow cone at angles close to the normal direction.
Below the threshold, spontaneous emission dominates.
The photon number is a maximum in the dark mode at the edge of the band gap and slowly decreases when the mode wavelength shifts to the allowed band.
Above the threshold, the lasing takes place in the bright modes which have nonzero Bloch wavevectors.
In this case, above the threshold the radiation pattern has a maximum at a nonzero angle.
Since the mode cooperation shifts the lasing from the dark mode to the bright modes, this widens the radiation pattern, see Fig.~\ref{fig7}b.

We indicated above that in~\cite{TennerACSPhot} it was demonstrated that the measured radiation pattern of a plasmonic DFB laser is wider than the one which is expected from the standard distributed feedback (DFB) theory for one-dimensional systems with finite size~\cite{Kogelnik1972DFB}.
Our numerical simulations show that this distinction arises owing to mode cooperation, something which does not appear in the one-dimensional case.

\section*{acknowledgement}
We thank Yu. E. Lozovik for helpful discussions. 

\clearpage

\begin{widetext}
\section*{Supplementary materials}
\section*{ Laser model}
The dynamics of the electromagnetic modes of a cavity and atoms in the Markov approximation is described by the master equation in the Lindblad form~\cite{CarmichaelOpen}
\begin{equation}
\frac{\partial }{{\partial t}}\hat \rho  =  - \frac{i}{\hbar }\left[ {{{\hat H}_a} + {{\hat H}_f} + \hat V,\hat \rho } \right] + {\hat L_a}\left[ {\hat \rho } \right] + \hat L_\sigma ^e\left[ {\hat \rho } \right] + \hat L_\sigma ^{ph}\left[ {\hat \rho } \right] + \hat L_\sigma ^{pump}\left[ {\hat \rho } \right] 
\tag{1}
\end{equation}
where ${\hat H_a} = \sum\limits_j {\hbar {\omega _j}\hat a_j^ + {{\hat a}_j}} $  is the Hamiltonian of the modes of the electromagnetic field, ${\hat H_a} = \sum\limits_m {\hbar {\omega _\sigma }\hat \sigma _m^ + {{\hat \sigma }_m}} $  is the Hamiltonian of the two-level atoms, and $\hat V = \hbar \sum\limits_{jm} {{\Omega _{jm}}\left( {\hat a_j^\dag {{\hat \sigma }_m} + \hat \sigma _m^\dag {{\hat a}_j}} \right)}$  is the Jaynes--Cummings Hamiltonian of interaction in the rotating-wave approximation.
Here, $\hat a_j^ + $  and ${\hat a_j}$  are the creation and annihilation operators of photons in the $j$th cavity mode, $\hat \sigma _m^ + $  and ${\hat \sigma _m}$  are the raising and lowering operators for the transition of the  $m$th two-level atom, ${\Omega _{jm}}$  is a coupling constant between the photons in the  $j$th cavity mode and the  $m$th atom (this is defined below), ${\omega _j}$  is the eigenfrequency of the  $j$th cavity mode, and ${\omega _\sigma }$  is the atom's transient frequency.
The term ${\hat L_a}\left[ {\hat \rho } \right] = \sum\limits_j {\frac{{{\gamma _j}}}{2}\left( {2{{\hat a}_j}\hat \rho \hat a_j^ +  - \hat a_j^ + {{\hat a}_j}\hat \rho  - \hat \rho \hat a_j^ + {{\hat a}_j}} \right)} $  describes the dissipation in the $j$th mode with dissipation rates ${\gamma _j}$, $\hat L_a^e\left[ {\hat \rho } \right] = \sum\limits_m {\frac{{{\gamma _D}}}{2}\left( {2{{\hat \sigma }_m}\hat \rho \hat \sigma _m^ +  - \hat \sigma _m^ + {{\hat \sigma }_m}\hat \rho  - \hat \rho \hat \sigma _m^ + {{\hat \sigma }_m}} \right)} $  and $\hat L_a^{ph}\left[ {\hat \rho } \right] = \sum\limits_m {\frac{{{\gamma _{ph} }}}{2}\left( {{{\hat D}_m}\hat \rho {{\hat D}_m} - \hat \rho } \right)}$  correspond to the energy and phase relaxation processes with rates ${\gamma _D}$  and ${\gamma _{ph} }$ , respectively, and  $\hat L_a^{pump}\left[ {\hat \rho } \right] = \sum\limits_m {\frac{{{\gamma _{pump}}}}{2}\left( {2\hat \sigma _m^ + \hat \rho {{\hat \sigma }_m} - {{\hat \sigma }_m}\hat \sigma _m^ + \hat \rho  - \hat \rho {{\hat \sigma }_m}\hat \sigma _m^ + } \right)} $ describes pumping of two-level atom with the rate ${\gamma _{pump}}$~\cite{CarmichaelOpen}.

Using the identity $\left\langle {\dot A} \right\rangle  = Tr\left( {\dot \rho A} \right)$  and the master equation (1), it is possible to derive a closed system of equations for the expectation values for the operators ${D_m} = \left\langle {\hat \sigma _m^ + {{\hat \sigma }_m} - {{\hat \sigma }_m}\hat \sigma _m^ + } \right\rangle $, ${\varphi _{jm}} = \left\langle { - i\hat a_j^ + {{\hat \sigma }_m}} \right\rangle $ and ${n_{jl}} = \left\langle {\hat a_j^ + {{\hat a}_l}} \right\rangle $.
To this end, we split the correlations between the average values of the number of photons and the population inversion, $\left\langle {{{\hat n}_{jk}}{{\hat D}_m}} \right\rangle  = \left\langle {{{\hat n}_{jk}}} \right\rangle \left\langle {{{\hat D}_m}} \right\rangle $~\cite{SiegmanLasers}, and neglect correlations between the raising and lowering operators of different atoms, $\left\langle {\hat \sigma _m^ + {{\hat \sigma }_l}} \right\rangle  = {\delta _{ml}}\left( {\left\langle {{{\hat D}_m}} \right\rangle  + 1} \right)/2$~\cite{HakenLaser}.
As a result, we obtain the following equations~\cite{Zyablovsky2017approach}:
\begin{equation}
\frac{{d{n_{jl}}}}{{dt}} =  - \left( {{\gamma _j} + {\gamma _l}} \right){n_{jl}} + i\left( {{\omega _j} - {\omega _l}} \right){n_{jl}} + \sum\limits_m {\left( {{\Omega _{lm}}{\varphi _{jm}} + \Omega _{jm}^*\varphi _{lm}^*} \right)} 
\tag{2}
\end{equation}
\begin{equation}
\frac{{d{D_m}}}{{dt}} =  - {\gamma _D}\left( {1 + {D_m}} \right) + {\gamma _{pump}}\left( {1 - {D_m}} \right) - 2\sum\limits_j {\left( {{\Omega _{jm}}{\varphi _{jm}} + \Omega _{jm}^*\varphi _{jm}^*} \right)}
\tag{3} 
\end{equation}
\begin{equation}
\frac{{d{\varphi _{jm}}}}{{dt}} =  -\gamma_{\sigma j} {\varphi _{jm}} + i\left( {{\omega _j} - {\omega _\sigma }} \right){\varphi _{jm}} + \frac{{\Omega _{jm}^*}}{2}\left( {{D_m} + 1} \right) + \sum\limits_l {\Omega _{lm}^*{n_{jl}}{D_m}} \tag{4}, 
\end{equation}
where $\gamma_{\sigma j} = \left( \gamma_{ph} + \gamma_j + (\gamma_D + \gamma_{pump})/2 \right)$.
In Eqs. (2)--(4), $D_m$  is the expectation value of the operator of the population inversion of the $m$th atom, and ${\varphi _{jm}}$  is the expectation value of the operator that describes the interaction between the electromagnetic field in the $j$th cavity mode and the $m$th atom.
${n_{jl}}$  is the expectation value of the operator of the number of photons in the $j$th cavity mode when $j = l$, and ${n_{jl}}$  is the expectation value of the operator that describes the transition of photons from the $l$th cavity mode to the $j$th cavity mode when $j \ne l$.
This operator arises from the interference between the electromagnetic field in the $j$th and $l$th cavity modes~\cite{Zyablovsky2017approach}.
In~\cite{Zyablovsky2017approach} it was demonstrated that neglecting the interference terms ${n_{jl}}$ results in instant propagation of the electromagnetic field.

\section*{Mode structure of two-dimensional plasmonic distributed-feedback laser}
In a two-dimensional plasmonic lattice, the electric and magnetic fields of the eigenmodes satisfy the Bloch condition.
The eigenmodes of the 2D plasmonic distributed-feedback laser can be calculated by the coupled-mode theory~\cite{ExterOptExp,TennerJOpt}.
Within this framework the electric field is expressed as
\begin{equation}
{\bf{E}}\left( {r,t} \right) = \left( {{E_x}\left( t \right){{\bf{u}}_x}\exp \left( {iGx} \right) + {E_{ - x}}\left( t \right){{\bf{u}}_{ - x}}\exp \left( { - iGx} \right) + {E_y}\left( t \right){{\bf{u}}_y}\exp \left( {iGy} \right) + {E_{ - y}}\left( t \right){{\bf{u}}_{ - y}}\exp \left( { - iGy} \right)} \right)\exp \left( {i{k_x}x + i{k_y}y} \right)\tag{5}
\end{equation}
Here, $G$  is the reciprocal lattice vector, ${{\bf{u}}_x}$,  ${{\bf{u}}_{ - x}}$, ${{\bf{u}}_y}$, and ${{\bf{u}}_{ - y}}$  are the eigenmodes of the system without holes.
${E_x}$ , ${E_{ - x}}$,  ${E_y}$, and ${E_{ - y}}$  are the amplitudes of the corresponding eigenmodes and ${k_{x,y}} = \omega /c\sin {\theta _{x,y}}$, where ${\theta _{x,y}}$ is the emission angle along the respective axes.

The amplitudes  ${E_x}$ , ${E_{ - x}}$,  ${E_y}$, and ${E_{ - y}}$ are found from
\begin{equation}
dE/dt = i\hat HE = i\left( {{{\hat H}_0} + \hat V} \right)E\tag{6}
\end{equation}
where 
\begin{equation}
{\hat H_0} = \left( {\begin{array}{*{20}{c}}
	{{\omega _0}}&\gamma &\kappa &\kappa \\
	\gamma &{{\omega _0}}&\kappa &\kappa \\
	\kappa &\kappa &{{\omega _0}}&\gamma \\
	\kappa &\kappa &\gamma &{{\omega _0}}
	\end{array}} \right) ,
\hat V\left( {{\theta _x},{\theta _y}} \right) = \left( {\begin{array}{*{20}{c}}
	{{c_2}\theta _x^2 + {c_1}{\theta _y}}&0&0&0\\
	0&{{c_2}\theta _x^2 - {c_1}{\theta _y}}&0&0\\
	0&0&{{c_1}{\theta _x} + {c_2}\theta _y^2}&0\\
	0&0&0&{ - {c_1}{\theta _x} + {c_2}\theta _y^2}
	\end{array}} \right)\tag{7}
\end{equation}
Here, ${\omega _0} = Gc/{n_{eff}}$ ($c$ is the speed of light), $\gamma $  and $\kappa$  are scattering amplitudes, ${c_1} = {\omega _0}/{n_{eff}}$, ${c_2} = {\omega _0}/\left( {2n_{eff}^2} \right)$, and ${n_{eff}}$  is an effective refractive index~\cite{ExterOptExp,TennerJOpt}.
The eigenvectors of $\hat H$  estimate the amplitudes ${E_x}$ , ${E_{ - x}}$,  ${E_y}$, and ${E_{ - y}}$.
The values of $\gamma $, $\kappa$, $c_1$, $c_2$ and $n_{eff}$ are taken from~\cite{TennerJOpt}.
Using them, we obtain the dispersion curve for the 2D distributed-feedback laser, see Figure 2 in the manuscript.

\section*{Calculation of the coupling constant}
The coupling constant of the electromagnetic field with the atom of the active medium can be expressed as
\begin{equation}
\hbar {\Omega _{im}} =  - {{\bf{d}}_m} \cdot {{\bf{E}}_i}({{\bf{r}}_m})\tag{8}
\end{equation}
where ${{\bf{d}}_m}$  is the dipole moment of the $m$th atom at the transition frequency and ${{\bf{E}}_i}({{\bf{r}}_m})$  is the electric field `per one quantum' in the $i$th mode at the position of the $m$th atom.
To normalize the electric field ${{\bf{E}}_i}({{\bf{r}}_m})$, we equate the energy of one quantum to the energy of the electric field in the resonator:
\begin{equation}
\hbar {\omega _i} = \frac{1}{{8\pi }}\int\limits_V {\left( {{{\left. {\frac{{\partial {\mathop{\rm Re}\nolimits} \left( {\varepsilon \omega } \right)}}{{\partial \omega }}} \right|}_{\omega  = {\omega _i}}}{{\left| {{{\bf{E}}_i}\left( {\bf{r}} \right)} \right|}^2} + {{\left| {{{\bf{H}}_i}\left( {\bf{r}} \right)} \right|}^2}} \right){d^3}{\bf{r}}} \tag{9}
\end{equation}
where $\varepsilon $  is the dielectric permittivity of the medium.

In a two-dimensional plasmonic lattice whose surface is parallel to the $xy$-plane, the electric and magnetic fields satisfy the Bloch condition,
\begin{equation}
{{\bf{E}}_i}\left( {\bf{r}} \right) = {E_i}\exp \left( {i{k_{xi}}x + i{k_{yi}}y} \right){{\bf{e}}_i}\left( {x,y,z} \right) \tag{10}
\end{equation}
\begin{equation}
{{\bf{H}}_i}\left( {\bf{r}} \right) = {H_i}\exp \left( {i{k_{xi}}x + i{k_{yi}}y} \right){{\bf{h}}_i}\left( {x,y,z} \right), \tag{11}
\end{equation}
where ${{\bf{e}}_i}\left( {x,y,z} \right) = {{\bf{e}}_i}\left( {x + {L_x},y + {L_y},z} \right)$  and ${{\bf{h}}_i}\left( {x,y,z} \right) = {{\bf{h}}_i}\left( {x + {L_x},y + {L_y},z} \right)$  are periodic functions with plasmonic lattice periods.
${L_x}$  and ${L_y}$  are the periods of the plasmonic lattice along the $x$-axis and the $y$-axis, respectively.
Below we will assume that ${{\bf{e}}_i}\left( {x,y,z} \right)$ and ${{\bf{h}}_i}\left( {x,y,z} \right)$  satisfy the normalization condition
\begin{equation}
\int\limits_{{V_G}} {{{\left| {{{\bf{e}}_i}\left( {\bf{r}} \right)} \right|}^2}{d^3}{\bf{r}}}  = \int\limits_{{V_G}} {{{\left| {{{\bf{h}}_i}\left( {\bf{r}} \right)} \right|}^2}{d^3}{\bf{r}}}  = {V_G} \tag{12}
\end{equation}
where ${V_G}$  is the volume of the active medium in one cell of the plasmonic lattice.

Suppose that the area under investigation contains an integral number $N$ of plasmonic lattice cells.
Then Eq. (9) may be rewritten as
\begin{equation}
\hbar {\omega _i} = \frac{N}{{8\pi }}\int\limits_{{V_{cell}}} {\left( {{{\left. {\frac{{\partial {\mathop{\rm Re}\nolimits} \left( {\varepsilon \omega } \right)}}{{\partial \omega }}} \right|}_{\omega  = {\omega _i}}}{{\left| {{E_i}} \right|}^2}{{\left| {{{\bf{e}}_i}\left( {\bf{r}} \right)} \right|}^2} + {{\left| {{H_i}} \right|}^2}{{\left| {{{\bf{h}}_i}\left( {\bf{r}} \right)} \right|}^2}} \right){d^3}{\bf{r}}} \tag{13}
\end{equation}
where ${V_{cell}}$  is the volume of the cell of the plasmonic lattice.

To calculate the coupling constant ${\Omega _{im}}$  between the field and amplifying medium, we introduce the parameter ${\eta _i}$.
It is determined by the ratio of the energy of the electric field of the $i$th eigenmode in the volume of the amplifying medium to the total energy of this mode defined by Eq. (9):
\begin{equation}
{\eta _i} = \frac{1}{{8\pi \hbar {\omega _i}}}\int\limits_{{V_G}} {{{\left. {\frac{{\partial {\mathop{\rm Re}\nolimits} \left( {\varepsilon \omega } \right)}}{{\partial \omega }}} \right|}_{\omega  = {\omega _i}}}{{\left| {{E_i}} \right|}^2}{{\left| {{{\bf{e}}_i}\left( {\bf{r}} \right)} \right|}^2}{d^3}{\bf{r}}} \tag{14}
\end{equation}
Assuming that ${\left. {\partial {\mathop{\rm Re}\nolimits} \left( {\varepsilon ({\bf{r}})\omega } \right)/\partial \omega } \right|_{\omega  = {\omega _i}}}$  is constant in the active medium, we obtain
\begin{equation}
{\eta _i} = \frac{1}{{8\pi \hbar {\omega _i}}}{V_G}{\left. {\frac{{\partial {\mathop{\rm Re}\nolimits} \left( {{\varepsilon _G}\omega } \right)}}{{\partial \omega }}} \right|_{\omega  = {\omega _i}}}{\left| {{E_i}} \right|^2} \tag{15}
\end{equation}
where ${\varepsilon _G}$  is the dielectric permittivity of the active medium.

From Eq. (15) we express the average amplitude of the electric field in the active medium via  ${\eta _i}$
\begin{equation}
\left| {{E_i}} \right| = \sqrt {\frac{{8\pi \,{\eta _i}\hbar {\omega _i}}}{{{V_G}{{\left. {\frac{{\partial {\mathop{\rm Re}\nolimits} \left( {{\varepsilon _G}\omega } \right)}}{{\partial \omega }}} \right|}_{\omega  = {\omega _i}}}}}} \tag{16}
\end{equation}
Using Eq. (16) we can find ${\Omega _{im}}$:
\begin{equation}
{\Omega _{im}} = \sqrt {\frac{{8\pi \,{\eta _i}{\omega _i}}}{{\hbar {V_G}{{\left. {\frac{{\partial {\mathop{\rm Re}\nolimits} \left( {{\varepsilon _G}\omega } \right)}}{{\partial \omega }}} \right|}_{\omega  = {\omega _i}}}}}} \left( {{{\bf{d}}_m} \cdot {{\bf{e}}_i}\left( {{{\bf{r}}_m}} \right)} \right)\exp \left( {i{k_{xi}}{x_m} + i{k_{yi}}{y_m}} \right) \tag{17}
\end{equation}
Eq. (17) for the coupling constant may be rewritten as
\begin{equation}
{\Omega _{im}} = {\Omega _{0im}}\exp \left( {i{k_{xi}}{x_m} + i{k_{yi}}{y_m}} \right){f_i}\left( {{{\bf{r}}_m}} \right) \tag{18}
\end{equation}
where ${f_j}\left( {{{\bf{r}}_m}} \right)$ is a periodic function with plasmonic lattice cell period which satisfies the normalization condition
\begin{equation}
\int\limits_{{V_G}} {{{\left| {{f_i}\left( {{{\bf{r}}_m}} \right)} \right|}^2}dV}  = {V_G} \tag{19}
\end{equation}
and
\begin{equation}
{\Omega _{0im}} = \sqrt {\frac{{8\pi {\eta _i}{\omega _i}{{\left| {{{\bf{d}}_m}} \right|}^2}}}{{\hbar {V_G}{{\left. {\frac{{\partial {\mathop{\rm Re}\nolimits} \left( {{\varepsilon _G}\omega } \right)}}{{\partial \omega }}} \right|}_{\omega  = {\omega _i}}}}}} \tag{20}
\end{equation}
For different allowed bands of the plasmonic lattice the functions ${f_i}\left( {{{\bf{r}}_m}} \right)$  are orthogonal to each other.

\section*{Numerical simulation}
The number of equations (2)--(4) increases linearly with the number of atoms and increases quadraticly with the number of laser modes.
As a consequence, the simulation of a real 2D plasmonic distributed-feedback laser requires enormous computational resources because the number of atoms in the active medium $N_{at} \sim 10^{11}$.

Note that the coupling constant (18) has two different spatial scales.
The small one is the period  of the plasmonic lattice $L \sim 500$\ nm.
This is the characteristic scale for the function  ${f_i}\left( {{{\bf{r}}_m}} \right)$.
The large scale is the Bloch wavevector.
Because we consider the modes near the forbidden gap, the maximal Bloch wavevector may be evaluated as ${\bf{k}} \sim 1/A \sim (100 \mu$m$)^{-1}$ where $A$ is the size of the plasmonic lattice.
Thus, we can perform an averaging over atoms which are placed in a region which is both larger than the lattice period and smaller than the inverse value of  the maximal Bloch wavevector.
For numerical simulation, we perform an averaging over one period of the plasmonic lattice.
Such averaging enables one to move to one effective atom per one cell, and decrease the total number of atoms to $10^3$.

The averaging procedure is as follows.
First, we divide the active medium into $M$ subwavelength areas with volume $V_{cell}$ with constant atomic concentration $n_c$.
Then we perform an averaging over each area by using the equality $\mathop \Sigma \limits_m  = \mathop \Sigma \limits_M {n_c}\mathop \smallint \limits_{{V_{cell}}} $.
As a result, from Eqs. (2)--(3) we obtain
\begin{equation}
\frac{{d{n_{jl}}}}{{dt}} =  - \left( {{\gamma _j} + {\gamma _l}} \right){n_{jl}} + i\left( {{\omega _j} - {\omega _l}} \right){n_{jl}} + \sum\limits_M {\left( {{n_c}\int\limits_{{V_{cell}}} {\left( {{\Omega _{lm}}{\varphi _{jm}} + \Omega _{jm}^*\varphi _{lm}^*} \right)dV} } \right)} \tag{21}
\end{equation}
\begin{equation}
\frac{d}{{dt}}\left( {\int\limits_{{V_{cell}}} {{D_m}dV} } \right) =  - {\gamma _D}\left( {{V_{cell}} + \int\limits_{{V_{cell}}} {{D_m}dV} } \right) + {\gamma _{pump}}\left( {{V_{cell}} - \int\limits_{{V_{cell}}} {{D_m}dV} } \right) - 2\sum\limits_j {\left( {\int\limits_{{V_{cell}}} {\left( {{\Omega _{jm}}{\varphi _{jm}} + \Omega _{jm}^*\varphi _{jm}^*} \right)dV} } \right)}  \tag{22}
\end{equation}
In Eq. (22) we divided all terms by ${n_c}$.
Multiplying Eq. (4) by ${\Omega _{jm}}$, and then carrying out the atom averaging in the area concerned, we obtain
\begin{equation}
\begin{array}{l}
\frac{d}{{dt}}\left( {\int\limits_{{V_{cell}}} {{\Omega _{jm}}{\varphi _{jm}}dV} } \right) =  - \left( { \gamma_{\sigma j}  - i\left( {{\omega _j} - {\omega _\sigma }} \right)} \right)\int\limits_{{V_{cell}}} {{\Omega _{jm}}{\varphi _{jm}}dV}  + \int\limits_{{V_{cell}}} {\frac{{{{\left| {{\Omega _{jm}}} \right|}^2}}}{2}\left( {{D_m} + 1} \right)dV}\\
+ \sum\limits_l {\int\limits_{{V_{cell}}} {\left( {{\Omega _{jm}}\Omega _{lm}^*{n_{jl}}{D_m}} \right)dV} } 
\end{array}  \tag{23}
\end{equation}
In Eq. (23) we divided all terms by ${n_c}$.

Let us introduce the variables ${\tilde \varphi _{jm}} = \frac{1}{{{V_{cell}}}}\int\limits_{{V_{cell}}} {{f_j}\left( {{{\bf{r}}_m}} \right){\varphi _{jm}}dV} $.
Assuming that $\exp \left( {i{{\bf{k}}_j}{{\bf{r}}}} \right)$ weakly changes in the averaging area,  we arrive at the following equations for the variables ${\tilde \varphi _{jm}}$:
\begin{equation}
\begin{array}{l}
\frac{d}{{dt}}{{\tilde \varphi }_{jm}} =  - \left( { \gamma_{\sigma j}  - i\left( {{\omega _j} - {\omega _\sigma }} \right)} \right){{\tilde \varphi }_{jm}} + \Omega _{0jm}^*\exp \left( { - i{{\bf{k}}_j}{{\bf{r}}_m}} \right)\frac{1}{{{V_{cell}}}}\int\limits_{{V_{cell}}} {\frac{{{{\left| {{f_j}\left( {{{\bf{r}}_m}} \right)} \right|}^2}}}{2}\left( {{D_m} + 1} \right)dV}  + \\
+ \sum\limits_l {\Omega _{0jm}^*\exp \left( { - i{{\bf{k}}_l}{{\bf{r}}_m}} \right){n_{jl}}\frac{1}{{{V_{cell}}}}\int\limits_{{V_{cell}}} {\left( {{f_j}\left( {{{\bf{r}}_m}} \right)f_l^*\left( {{{\bf{r}}_m}} \right){D_m}} \right)dV} } 
\end{array}  \tag{24}
\end{equation}
If ${D_m}$  weakly changes in the averaging area, then Eq. (24) takes the form
\begin{equation}
\frac{{d{{\tilde \varphi }_{jm}}}}{{dt}} =   - \left( { \gamma_{\sigma j}  - i\left( {{\omega _j} - {\omega _\sigma }} \right)} \right){\tilde \varphi _{jm}} + \tilde \Omega _{jm}^*{\beta _{jm}}\frac{{\left( {{D_m} + 1} \right)}}{2} + \sum\limits_l {{\eta _{jl}}{\beta _{jm}}\tilde \Omega _{lm}^*{n_{jl}}{D_m}}   \tag{25}
\end{equation}
where
\begin{equation}
{\tilde \Omega _{jm}} = {\Omega _{0jm}}\exp \left( {i{{\bf{k}}_j}{{\bf{r}}}} \right) \tag{26}
\end{equation}
\begin{equation}
{\beta _{jm}} = \frac{{\int\limits_{{V_{cell}}} {{{\left| {{f_j}\left( {{{\bf{r}}_m}} \right)} \right|}^2}dV} }}{{{V_{cell}}}}  \tag{27}
\end{equation}
and
\begin{equation}
{\eta _{jl}} = \frac{{\int\limits_{{V_{cell}}} {\left( {{f_j}\left( {{{\bf{r}}}} \right)f_l^*\left( {{{\bf{r}}}} \right)} \right)dV} }}{{\int\limits_{{V_{cell}}} {{{\left| {{f_j}\left( {{{\bf{r}}}} \right)} \right|}^2}dV} }}  \tag{28}
\end{equation}
If the modes are related to one allowed band, then ${\eta _{jl}} \approx 1$, if the modes are related to different allowed bands, then ${\eta _{jl}} \approx 0$.
Note that when the averaging area is one cell of the plasmonic lattice, then due to the normalization condition (19), we have ${\beta _{jm}} = 1$.
If ${D_m}$  weakly changes in the averaging area, then Eq. (22) may be rewritten as
\begin{equation}
\frac{{d{D_m}}}{{dt}} =  - {\gamma _D}\left( {1 + {D_m}} \right) + {\gamma _{pump}}\left( {1 - {D_m}} \right) - 2\sum\limits_j {\left( {{{\tilde \Omega }_{jm}}{{\tilde \varphi }_{jm}} + \tilde \Omega _{jm}^*\tilde \varphi _{jm}^*} \right)}  \tag{29}
\end{equation}
Eq. (21) for the variables ${n_{jk}}$  contains the terms $\int\limits_{{V_{cell}}} {{\Omega _{km}}{\varphi _{jm}}dV}$.
The equation for these variables is
\begin{equation}
\begin{array}{l}
\frac{d}{{dt}}\left( {\int\limits_{{V_{cell}}} {{\Omega _{km}}{\varphi _{jm}}dV} } \right) =   - \left( { \gamma_{\sigma j}  - i\left( {{\omega _j} - {\omega _\sigma }} \right)} \right)\int\limits_{{V_{cell}}} {{\Omega _{km}}{\varphi _{jm}}dV}  + \int\limits_{{V_{cell}}} {\frac{{{\Omega _{km}}\Omega _{jm}^*}}{2}\left( {{D_m} + 1} \right)dV}  + \\
+ \sum\limits_l {\int\limits_{{V_{cell}}} {\left( {{\Omega _{km}}\Omega _{lm}^*{n_{jl}}{D_m}} \right)dV} } 
\end{array}  \tag{30}
\end{equation}
Let us set the new variable ${\tilde \varphi _{kjm}} = \frac{1}{{{V_{cell}}}}\int\limits_{{V_{cell}}} {{f_k}\left( {{{\bf{r}}_m}} \right){\varphi _{jm}}dV}$.
If $\exp \left( {i{{\bf{k}}_j}{{\bf{r}}_m}} \right)$  weakly changes in the averaging area, then ${\tilde \varphi _{kjm}}$ satisfies the equation
\begin{equation}
\begin{array}{l}
\frac{d}{{dt}}{{\tilde \varphi }_{kjm}} =  - \left( { \gamma_{\sigma j}  - i\left( {{\omega _j} - {\omega _\sigma }} \right)} \right){{\tilde \varphi }_{kjm}} + \Omega _{0jm}^*\exp \left( { - i{{\bf{k}}_j}{{\bf{r}}_m}} \right)\frac{1}{{{V_{cell}}}}\int\limits_{{V_{cell}}} {\frac{{{f_k}\left( {{{\bf{r}}_m}} \right)f_j^*\left( {{{\bf{r}}_m}} \right)}}{2}\left( {{D_m} + 1} \right)dV}  + \\
+ \sum\limits_l {\Omega _{0jm}^*\exp \left( { - i{{\bf{k}}_l}{{\bf{r}}_m}} \right){n_{jl}}\frac{1}{{{V_{cell}}}}\int\limits_{{V_{cell}}} {\left( {{f_k}\left( {{{\bf{r}}_m}} \right)f_l^*\left( {{{\bf{r}}_m}} \right){D_m}} \right)dV} } 
\end{array}  \tag{31}
\end{equation}
If ${D_m}$ weakly changes in the averaging area as well, then Eq. (31) takes the form
\begin{equation}
\frac{d}{{dt}}{\tilde \varphi _{kjm}} =   - \left( { \gamma_{\sigma j}  - i\left( {{\omega _j} - {\omega _\sigma }} \right)} \right){\tilde \varphi _{kjm}} + \tilde \Omega _{jm}^*{\eta _{kj}}{\beta _{km}}\frac{{\left( {{D_m} + 1} \right)}}{2} + \sum\limits_l {\tilde \Omega _{lm}^*{\eta _{kl}}{\beta _{km}}{n_{jl}}{D_m}}   \tag{32}
\end{equation}

Eq. (21) for ${n_{jk}}$  may be rewritten as
\begin{equation}
\frac{{d{n_{jk}}}}{{dt}} =  - \left( {{\gamma _j} + {\gamma _k}} \right){n_{jk}} + i\left( {{\omega _j} - {\omega _k}} \right){n_{jk}} + {N_a}\sum\limits_M {\left( {{{\tilde \Omega }_{km}}{{\tilde \varphi }_{kjm}} + \tilde \Omega _{jm}^*\tilde \varphi _{jkm}^*} \right)} \tag{33}
\end{equation}
Here, ${N_a}$  is the number of atoms in the averaging area.

If the  $k$th and  $j$th modes lie in one allowed band, then ${\eta _{kl}} \approx {\eta _{jl}}$, ${\beta _{km}} \approx {\beta _{jm}}$  and ${\eta _{kj}} \approx 1$. 
In this case we obtain that
\begin{equation}
\frac{d}{{dt}}{\tilde \varphi _{kjm}} \approx   - \left( { \gamma_{\sigma j}  - i\left( {{\omega _j} - {\omega _\sigma }} \right)} \right){\tilde \varphi _{kjm}} + \tilde \Omega _{jm}^*{\beta _{jm}}\frac{{\left( {{D_m} + 1} \right)}}{2} + {\beta _{jm}}\sum\limits_l {\tilde \Omega _{lm}^*{\eta _{jl}}{n_{jl}}} {D_m} = \frac{{d{{\tilde \varphi }_{jm}}}}{{dt}} \tag{34}
\end{equation}
If  the $k$th and  $j$th modes lie in different allowed bands, then ${\eta _{kj}} \approx 0$ and we obtain that
\begin{equation}
\frac{d}{{dt}}{\tilde \varphi _{kjm}} =   - \left( { \gamma_{\sigma j}  - i\left( {{\omega _j} - {\omega _\sigma }} \right)} \right){\tilde \varphi _{kjm}} + {\beta _{km}}\sum\limits_l {\tilde \Omega _{lm}^*{\eta _{kl}}{n_{jl}}{D_m}}  \tag{35}
\end{equation}
where the sum is only for modes that lie in the one allowed band with  the $k$th mode.

Thus, to find the population inversion and the number of photons, it is not necessary to know the values of ${n_{jk}}$  for those $k$th and  $j$th modes which relate to different allowed bands.
The final equations have the form
\begin{equation}
\frac{{d{n_{jl}}}}{{dt}} =  - \left( {{\gamma _j} + {\gamma _l}} \right){n_{jl}} + i\left( {{\omega _j} - {\omega _l}} \right){n_{jl}} + {N_a}\sum\limits_m {\left( {{{\tilde \Omega }_{lm}}{{\tilde \varphi }_{jm}} + \tilde \Omega _{jm}^*\tilde \varphi _{lm}^*} \right)} \tag{36}
\end{equation}
\begin{equation}
\frac{{d{D_m}}}{{dt}} =  - {\gamma _D}\left( {1 + {D_m}} \right) + {\gamma _{pump}}\left( {1 - {D_m}} \right) - 2\sum\limits_j {\left( {{{\tilde \Omega }_{jm}}{{\tilde \varphi }_{jm}} + \tilde \Omega _{jm}^*\tilde \varphi _{jm}^*} \right)} \tag{37}
\end{equation}
\begin{equation}
\frac{d}{{dt}}{\tilde \varphi _{jm}} =  - \left( {{\gamma _{\sigma j}} - i\left( {{\omega _j} - {\omega _\sigma }} \right)} \right){\tilde \varphi _{jm}} + \tilde \Omega _{jm}^*{\beta _{jm}}\frac{{\left( {{D_m} + 1} \right)}}{2} + {\beta _{jm}}\left( {\sum\limits_l {{\eta _{jl}}\tilde \Omega _{lm}^*{n_{jl}}} } \right){D_m} \tag{38}
\end{equation}
As we noted above, the parameter ${\eta _{jl}} \approx 1$  for modes of one allowed band and ${\eta _{jl}} \ll  1$  for modes of different allowed bands.
For numerical simulation, we use ${\eta _{jl}} = 1$  for modes of one allowed band and ${\eta _{jl}} = 0$  for modes of different allowed bands.
In other words, we take into account interference terms only for modes which lie in the same allowed band.

Usually in plasmonic DFB lasers with a dye or semiconductor active medium, the following relation holds between the dissipation rates, ${\gamma _{ph} } \gg {\gamma _j},{\gamma _D},{\gamma _{pump}}$.
In this case it is possible to adiabatically eliminate the variables ${\tilde \varphi _{jm}}$:
\begin{equation}
{\tilde \varphi _{jm}} = {\beta _{jm}}\frac{{\tilde \Omega _{jm}^*\left( {{D_m} + 1} \right) + 2\sum\limits_k {\tilde \Omega _{km}^*{n_{jk}}{D_m}} }}{{2{\Delta _j}}} \tag{39}
\end{equation} 
and rewrite Eqs. (36)--(38) in the form
\begin{equation}
\frac{{d{n_{jj}}}}{{dt}} =  - 2{\gamma _j}{n_{jj}} + {N_{a}}{\beta _{jm}}\sum\limits_m {\frac{{{{\left| {{{\tilde \Omega }_{jm}}} \right|}^2}{\gamma _{\sigma j}}}}{{{{\left| {{\Delta _j}} \right|}^2}}}} \left( {2{n_{jj}}{D_m} + {D_m} + 1} \right) + {N_{a}}{\beta _{jm}}\sum\limits_{m,k,k \ne j} {\left( {\frac{{{{\tilde \Omega }_{jm}}\tilde \Omega _{km}^*{n_{jk}}}}{{{\Delta _j}}}{\rm{ + c}}.{\rm{c}}.} \right){D_m}} \tag{40}
\end{equation}
\begin{equation}
\begin{array}{l}
\frac{{d{n_{jl}}}}{{dt}} =  - \left( {{\gamma _j} + {\gamma _l}} \right){n_{jl}} + i\left( {{\omega _j} - {\omega _l}} \right){n_{jl}} + {N_{a}}\sum\limits_m {\left( {\frac{{{{\tilde \Omega }_{lm}}\tilde \Omega _{jm}^*\left( {{D_m} + 1} \right)\left( {{\Delta _j} + \Delta _l^*} \right)}}{{2{\Delta _j}\Delta _l^*}}} \right)}  + \\
\nonumber +{N_{a}}\sum\limits_{m,k,k \ne j} {\left( {\frac{{{\beta _{jm}}{{\tilde \Omega }_{lm}}\tilde \Omega _{km}^*{n_{jk}}}}{{{\Delta _j}}} + \frac{{\beta _{lm} {{\tilde \Omega }_{km}}\tilde \Omega _{jm}^*{n_{lk}}}}{{\Delta _l^*}}} \right){D_m}}
\end{array} \tag{41}
\end{equation}
\begin{equation}
\begin{array}{l}
\frac{{d{D_m}}}{{dt}} =  - {\gamma _D}\left( {1 + {D_m}} \right) + {\gamma _{pump}}\left( {1 - {D_m}} \right) - 2\sum\limits_j {\frac{{{\beta _{jm}}{{\left| {{{\tilde \Omega }_{jm}}} \right|}^2}{\gamma _{\sigma j}}}}{{{{\left| {{\Delta _j}} \right|}^2}}}} \left( {2{n_{jj}}{D_m} + {D_m} + 1} \right) - \\
\nonumber - 2\sum\limits_{j,k,k \ne j} {\left( {\frac{{{\beta _{jm}}{{\tilde \Omega }_{jm}}\tilde \Omega _{km}^*{n_{jk}}}}{{{\Delta _j}}} + {\rm{c}}.{\rm{c}}.} \right){D_m}} 
\end{array}\tag{42}
\end{equation}
where ${\Delta _j} = {\gamma _{\sigma j}} - i\left( {{\omega _j} - {\omega _\sigma }} \right)$.

\begin{figure*}
	\centering
	\includegraphics[width=0.95\columnwidth]{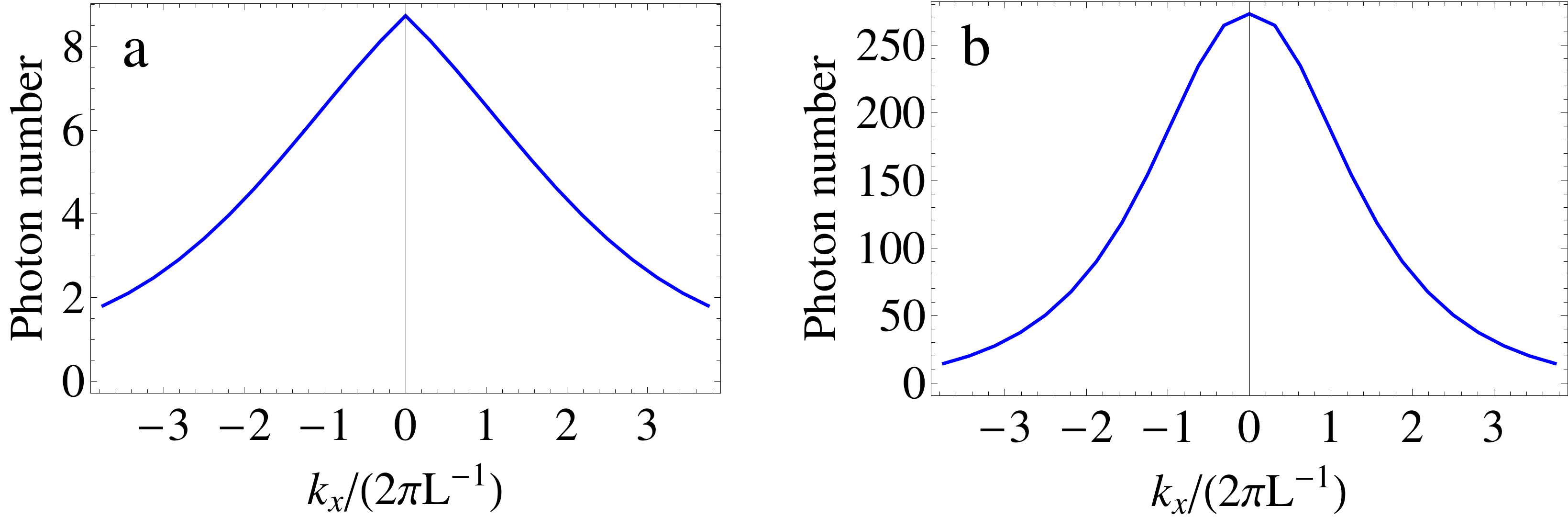}\\
	\caption{The mode distribution of the photon number of one-dimensional plasmonic DFB laser below (a) and above (b) generation threshold.}\label{fig1S}
\end{figure*}
\begin{figure*}
	\centering
	\includegraphics[width=0.95\columnwidth]{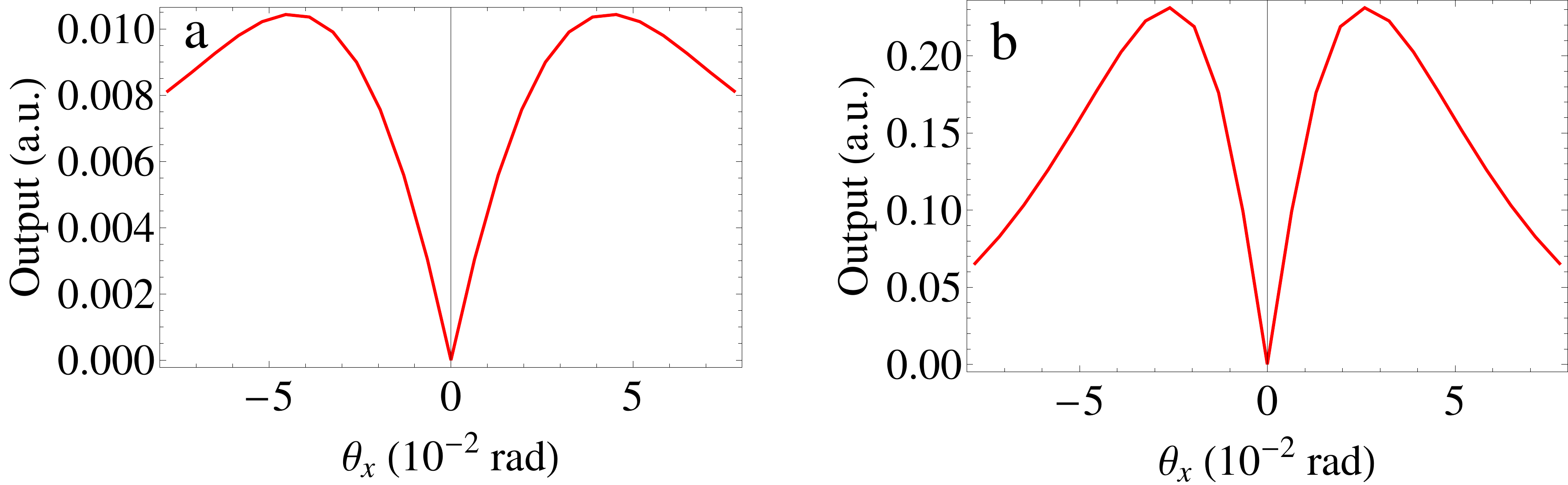}\\
	\caption{The radiation pattern of one-dimensional plasmonic DFB laser below (a) and above (b) generation threshold.}\label{fig2S}
\end{figure*}

Eqs. (40)--(42) describe the long-time dynamics of Eqs. (36)--(38).
In our paper we restrict ourselves to studying the operation of the 2D plasmonic DFB laser near stationary regimes.
For this reason, we use Eqs. (40)--(42) instead of Eqs. (36)--(38).

Eqs. (40)--(42) take into account the nonuniform distribution of the electromagnetic field and atomic polarization and population inversion in a plasmonic lattice.
They describe both the spontaneous emission of each atom to each mode and interference between electromagnetic fields from different modes.
The last two points are of great importance for DFB lasers.

\section*{The mode distribution of the photon number and the radiation pattern of one-dimensional plasmonic DFB laser}

In the main text, we apply the developed model, Eqs. (40)--(42), for a two-dimensional plasmonic DFB laser. Here, we present the results for a one-dimensional one. In Fig. 1, the photon number in the mode with the Bloch wavevector $k_{x}$ is shown. It is seen that the maximum of the photon number is in the mode with zero wavevector at the edge of the band gap for both cases, below and above threshold. In contrast, in the two-dimensional plasmonic DFB laser above the threshold, the maximum of the photon number is in the mode with non-zero Bloch wavevector, see Fig.4b in the main text.

In Fig. 2, the radiation pattern of the one-dimensional plasmonic laser is shown. The radiation pattern above the generation threshold coincides with prediction of Kogelnik theory ~\cite{TennerJOpt}. However, in the two-dimensional case, more accurate theory, which we presented here, as well as experimental data ~\cite{TennerJOpt} demonstrate a wider radiation pattern, see Fig. 7b in the main text.

\end{widetext}

\bibliography{ref1}

\end{document}